\documentclass[%
 reprint,
superscriptaddress,
 amsmath,amssymb,
 aps,
]{revtex4-2}

\usepackage{graphicx}% Include figure files
\usepackage{dcolumn}% Align table columns on decimal point
\usepackage{bm}% bold math
\usepackage{amsmath}
\usepackage{booktabs}
\usepackage{amsfonts}
\usepackage{amsthm}
\usepackage{comment}
\usepackage{tabularx}
\usepackage{color,soul}
\usepackage{wrapfig}
\usepackage{float}

\DeclareMathOperator{\E}{\mathbb{E}}
\DeclareMathOperator{\V}{\mathbb{V}}

\DeclareMathOperator{\Cov}{Cov}
\begin{document}

\preprint{APS/123-QED}

\title{Uncertainty quantification and estimation in differential dynamic microscopy}% Force line breaks with \\

\author{Mengyang Gu}
   \thanks{Equal contribution}
 \affiliation{Department of Statistics and Applied Probability, University of California, Santa Barbara CA 93106, USA}%

\author{Yimin Luo}
\thanks{Equal contribution}
\affiliation{Department of Chemical Engineering, University of California, Santa Barbara CA 93106, USA}%
\affiliation{Department of Mechanical Engineering, University of California, Santa Barbara CA 93106, USA}%

\author{Yue He}
\affiliation{Department of Statistics and Applied Probability, University of California, Santa Barbara CA 93106, USA}%

\author{Matthew E. Helgeson}
\affiliation{Department of Chemical Engineering, University of California, Santa Barbara CA 93106, USA}%

\author{Megan T. Valentine}
 \email{Corresponding authors: mengyang@pstat.ucsb.edu; helgeson@ucsb.edu; valentine@engineering.ucsb.edu}
 \affiliation{Department of Mechanical Engineering, University of California, Santa Barbara CA 93106, USA}%

\date{\today}% It is always \today, today,

\begin{abstract}

Differential dynamic microscopy (DDM) is a form of video image analysis that combines the sensitivity of scattering and the direct visualization benefits of microscopy. DDM is broadly useful in determining dynamical properties including the intermediate scattering function for many spatiotemporally correlated systems. Despite its straightforward analysis, DDM has not been fully adopted as a routine characterization tool, largely due to computational cost and lack of algorithmic robustness.  We present statistical analysis that quantifies  the noise, reduces the computational order and enhances the robustness of DDM analysis. We propagate the image noise through the Fourier analysis, which allows us to comprehensively study the bias in different estimators of  model parameters, and we derive a different way to detect whether the bias is negligible. 
Furthermore, through use of Gaussian process regression (GPR), we find that predictive samples of the image structure function require only around 0.5\%--5\% of the Fourier transforms of the observed quantities. This vastly reduces computational cost, while preserving information of the quantities of interest, such as quantiles of the image scattering function, for subsequent analysis. The approach, which we call DDM with uncertainty quantification (DDM-UQ), is validated using both simulations and experiments with respect to accuracy and computational efficiency, as compared with conventional DDM and multiple particle tracking. Overall, we propose that DDM-UQ lays the foundation for important new applications of DDM, as well as to high-throughput characterization. We implement the fast computation tool in a new, publicly available MATLAB software package.

\end{abstract}
\maketitle
\setstcolor{blue}

\section{Introduction}

Microscopy has become an essential tool for probing dynamical processes in complex materials and systems, but typically requires sophisticated video image analysis to obtain quantitative information. Although real-space analysis methods retain information regarding individualistic processes within an image, feature tracking algorithms such as multiple particle tracking (MPT) are often computationally expensive and require user interactivity to determine algorithmic parameters to isolate the dynamical process(es) of interest \cite{crocker1996methods,savin2007statistical}. By contrast, Fourier transform-based analysis retains the statistical information encoded within the entire image, and is therefore more sensitive to low-signal processes as well as more robust to non-ideal imaging conditions and optically dense systems \cite{giavazzi2009scattering,giavazzi2014digital}. In this way, Fourier microscopy combines the advantages of real-space imaging in feature identification and segmentation with ensemble-level statistical precision of Fourier-space analysis.

Of the various Fourier-space based approaches available, differential dynamic microscopy (DDM) \cite{cerbino2008differential} has emerged as a powerful and versatile analysis method to quantify spatiotemporally correlated dynamics from video microscopy data. This versatility stems from its compatibility with a broad range of microscopy imaging modes, easy setup with instrumentation available in most research laboratories and straightforward analysis routines. DDM has been applied to study an ever-broadening range of phenomena in soft and biological matter systems \cite{giavazzi2014digital,giavazzi2009scattering,bayles2017probe}, including analysis of the dynamics of concentrated particle suspensions \cite{lu2012characterizing}, motions of swimming bacteria \cite{martinez2012differential}, binary mixture of molecular fluids \cite{giavazzi2016equilibrium}, and the coarsening dynamics of phase separating colloidal gels \cite{gao2015microdynamics}. While it is common to assume the material to be isotropic, anisotropic properties such as the viscoelasticity of nematic liquid crystals can also be extracted \cite{giavazzi2014viscoelasticity}.

For a more comprehensive overview of DDM and its various applications, the interested reader is referred to various reviews on the topic \cite{giavazzi2009scattering,giavazzi2014digital,cerbino2017perspective}. This work is concerned with the development of a comprehensive statistical framework that aims at quantifying errors, reducing computational cost and enhancing the robustness of the analysis of differential dynamic microscopy (DDM) data. Similar developments have been made previously for MPT analysis \cite{savin2005static,savin2007statistical}, and have greatly improved the robustness and algorithmic development of MPT in various applications. We therefore anticipate similar benefits from a more thorough investigation of uncertainty for DDM. To better motivate these developments, we first summarize the analysis procedure of DDM,  estimators employed to extract physical parameters,  and highlight the features and limitations of DDM that inspired this study.

In DDM, a time sequence of image stacks represented by the intensity matrix $I({\bf x},t)$ is processed using a Fourier-based technique where $\mathbf x=(x_1,x_2)$ denotes two spatial coordinates and $t \in [t_{min}, t_{max}]$. One first calculates the difference in intensity at each pixel location between two frames separated by a lag time $\Delta t$: 
\begin{equation}
\Delta I({\bf x},t,\Delta t) = I({\bf x},t+\Delta t)-I({\bf x},t).
\label{diff}
\end{equation}
The intensity differences are then Fourier transformed and the absolute values squared to obtain the normalized squared intensity function in Fourier space:
\begin{equation}
|\Delta \hat I({\bf q},t,\Delta t)|^2=|\mathcal F(\Delta I({\bf x},t,\Delta t))|^2,
\label{eq:squared intensity}
\end{equation}
where $\mathcal F(\cdot)$ denotes the operator of the 2D discrete Fourier transformation (DFT), ${\bf q} =  (q_1, q_2)$ is a coordinate wave vector in reciprocal space.

The ensemble average of Eq. (\ref{eq:squared intensity}) is computed to obtain the dynamic image structure function $D({\bf q},\Delta t)$:
\begin{align}
D({\bf q},\Delta t) &= \langle \vert \Delta \hat I({\bf q},t,\Delta t)\vert^2 \rangle \nonumber\\
&=\frac{1}{n_{\Delta t} n_{q}}\sum_{t \in \mathcal S_{\Delta t}}\sum_{(q_1,q_2)\in \mathcal S_q}  \vert \Delta \hat I({\bf q},t,\Delta t)\vert^2, 
\label{eq: nqnt}
\end{align}
 where  $\langle .\rangle$ denotes averaging across all instances of $t\in \mathcal S_{\Delta t} $ and $(q_1, q_2)\in \mathcal S_q$ with sets $\mathcal S_{\Delta t}=\{t: t_{min} \leq t \leq t_{max}-\Delta t \}$. The sizes of the two sets are denoted by $n_{\Delta t} =\# \mathcal S_{\Delta t}$ and $n_q =\# \mathcal S_q$, respectively. 

 We only consider isotropic materials in this work, in which case $D({\bf q},\Delta t) = D(q,\Delta t)$ where $\mathcal S_q=\{(q_1,q_2):q^2_1+q^2_2=q^2 \}$. However, the approach can be generalized to retain a multi-dimensional $q$-dependence as desired \cite{reufer2012differential}. 
 
The following representation is routinely used to relate observables commonly associated with scattering analysis to the observations of $D({q},\Delta t)$  \cite{cerbino2008differential,bayles2017probe,giavazzi2009scattering}: 
 \begin{equation}
D_o(q,\Delta t) = A({q})\left( 1-f({ q},\Delta t) \right) + B(q, \Delta t),
\label{equ:D_o_q_t}
\end{equation}
\noindent where $A({q})$ is determined by the properties of the imaged material and imaging optics,  $B(q,\Delta t)$ is determined by the noise of the detection chain, and the subscript `o' denotes the observed value.  As will be discussed in Sec. \ref{subsec:UQ}, the mean of $B(q,\Delta t)$ is a constant value shared across all $q$ and $\Delta t$ values, whereas the variance of $B(q, \Delta )$ depends on the values of $q$ and $\Delta t$.  The intermediate scattering function (ISF), $f(q, \Delta t)$, is in principle the same as that measured in conventional light scattering measurements such as dynamic light scattering (DLS). The ISF quantifies how the dynamic structure decorrelates over the observed length scale $1/q$ in Fourier space and timescale $\Delta t$ in real space, which encodes the physical dynamics of the observed system. In general, for randomly-fluctuating, ergodic systems, $f({ q}, \Delta t \rightarrow 0) = 1$ and $f({ q}, \Delta t \rightarrow \infty) = 0$, and thus, $ D_o(q,\Delta t \rightarrow 0) = B(q, \Delta t)$ and $D_o(q,\Delta t \rightarrow \infty) = A(q)+B(q,\Delta t)$. 

DDM's ultimate integration into the characterization workflows of a diverse range of systems is not without challenges. First, we note that, because DDM operates on a series of finite-exposure images taken of time-fluctuating processes, the measured DDM signal will contain inherent error related to both static and dynamic effects, much in the same way MPT incurs static and dynamic errors associated with the imaging process \cite{savin2005static}. Because of this, the observed value $D_o(q,\Delta t)$ will in general not be equal to the ``true'' image structure function $D(q,\Delta t)$ that would be obtained from an ideal imaging system. Separating the signal from the noise in estimating the image structure function requires properly quantifying the uncertainty of the background noise that propagates through the analysis \cite{cerbino2008differential}. 
 The observed intensity $D_o(q,\Delta t)$ typically overestimates $D(q,\Delta t)$, as the mean of the noise term $B(q,\Delta t)$ is positive, and is twice as large as the variance of the noise in the original images \cite{giavazzi2009scattering}. In this study, we show that it is critically important to obtain an accurate estimate of the mean of the noise term $B(q,\Delta t)$, which we denote as $B_{est}$, to extract dynamic information from systems, such as the mean squared displacement, using DDM.

Several distinct methods to obtain the noise estimator $B_{est}$ in DDM have been proposed in prior studies. However, to our knowledge, there has yet to be a detailed study of how the choice of estimator affects the estimation of dynamic properties. For instance, $B_{est}$ has been assumed to be 0 \cite{kurzthaler2018probing}, estimated by the minimum value of $D(q, \Delta t)$ at the temporal resolution $\Delta t_{min}$, denoted as $D_{min}(\Delta t_{min})$ \cite{bayles2017probe}, as the average of the high-$q$ limit of the observed image structure function $\langle D_o(q_{max},\Delta t)\rangle_{\Delta t}$ \cite{escobedo2018microliter, cerbino2017dark}, or as the ensemble average of the static power spectrum $ 2\langle\vert \hat{I}_o (q_{max}, t)\vert^2\rangle_t$ \cite{giavazzi2018tracking}. Although a particular estimator of the noise term may work well under certain experimental conditions, we will show that all estimators are biased in general (see Section \ref{subsec:UQ}), and ultimately introduce a way to discern whether the bias is negligible, or if additional measurements are needed to estimate the noise. Indeed, in both simulated scenarios and real experiments, we found that the bias of noise estimation can substantially  impact the estimation accuracy of system dynamics. Finally, we note that $B_{est}$ is sometimes treated as a fitting parameter and estimated along with other parameters in the model of $f(q,\Delta t)$
\cite{giavazzi2009scattering, moon2012two,lee2021myosin}. This approach is applied to analyzing the active actin dynamics from \cite{lee2021myosin} in Sec. \ref{subsec:active_actin}. In general, we find that estimation of the noise parameter is the most challenging among all  parameters, and could lead to a poor fit to the observed values. The first contribution of this work is a formal analysis of  error propagation and comparison of different estimators in representative experimental contexts.

After obtaining the estimate of the mean of the noise $B_{est}$, the amplitude parameter $A(q)$ may be estimated by the plateau of intensity through  $D_o(q,\Delta t \rightarrow \infty) = A(q)+B_{est}$  \cite{bayles2017probe,escobedo2018microliter}, or through the connection $A(q)+B_{est} =\langle\vert \hat{I}_o (q,t)\vert^2\rangle_t$ \cite{cerbino2008differential,giavazzi2018tracking,giavazzi2020multiple}.  In practice, it may be difficult to accurately obtain $D_o(q,\Delta t \rightarrow \infty)$ as there are very few observations available for intensities at large $\Delta t$. Thus, we find that estimation of $A(q)$ through $A(q)+B_{est} =\langle\vert \hat{I}_o (q,t)\vert^2\rangle_t$ is often more reliable, as it does not require the intensity to reach plateau at large $\Delta t$. These approaches will be compared using simulated and experimental observations.

A second challenge of DDM analysis is the computational cost: the most computationally expensive step for DDM is performing a 2D fast Fourier transformation for each pixel of the difference for each of the image pairs ($T\times (T-1)/2$ pairs, where $T\sim 10^3$--$10^4$ is the total number of time points). For $T$ images with size $N\times N$ pixels ($N \sim 10^2$--$10^3$), this requires $O(T^2 N^2\log_2(N))$ computational operations. Although there has been recent progress in accelerated computation that takes advantage of contemporary computational efficiency for Fourier-based image analysis \cite{norouzisadeh2020increased,lu2012characterizing}, these approaches still require resolving the DDM signal over the full sampled space of $q$ and $\Delta t$.  To overcome the computational challenge, we use a probabilistic approach to downsample the image stacks and reconstruct all image structure functions based on a fraction of observations, which dramatically reduces the computational cost of the required fast Fourier transform (FFT) algorithm.

The third, and often overlooked aspect, is the robustness of the algorithm(s) for decomposing $D_o(q,\Delta t)$ into its more physically meaningful components through Eq. (\ref{equ:D_o_q_t}). To ensure applicability to a wide range of materials, it is desirable to allow for an arbitrary form of $f(q,\Delta t)$, such that the method does not require prior knowledge of the system's dynamical properties. Moreover, the estimators used for $A(q)$ and $B(q)$ may contain bias, which can cause the algorithm to be less robust for both small and large $\Delta t$'s in estimating $f(q,\Delta t)$ and quantities derived from it \cite{bayles2017probe}.  Here we generate the predictive samples based on the observed image structure function $D_o(q,\Delta t)$ and use the predictive median to derive physical parameters within the systems; this approach is more robust than a simple ensemble based on $D_o(q,\Delta t)$ at selected wave vectors.

An exemplifying context for the potential advantages gained by overcoming these limitations is the recent application of DDM to passive probe microrheology as an alternative to conventional approaches such as MPT \cite{bayles2017probe,edera2017differential}. MPT-based passive probe microrheology involves imaging the Brownian fluctuations of embedded colloidal probes in order to resolve their mean square displacements (MSD) $\langle \Delta r^2(\Delta t)\rangle$ \cite{crocker1996methods}, which in the limit of homogeneous, uniform materials can be related to their linear viscoelastic moduli \cite{mason1997particle,Mason.PRL.1995}. DDM offers an alternative approach to MPT in estimating the MSD through the estimation of $f(q,\Delta t)$ using Eq. (\ref{equ:D_o_q_t}). DDM presents a number of advantages in this regard, including applicability and better statistical precision to low-signal or optically dense probes and materials \cite{bayles2017probe,cerbino2017dark,safari2015differential,sentjabrskaja2016anomalous}. Importantly, since DDM requires no \textit{a priori} user-input parameters associated with the probes or imaging system, it has the potential to provide automated, user-free analysis that could enable high-throughput characterization \cite{cerbino2017perspective}.

\begin{figure*}[t]
    \centering
    \includegraphics[width = \textwidth]{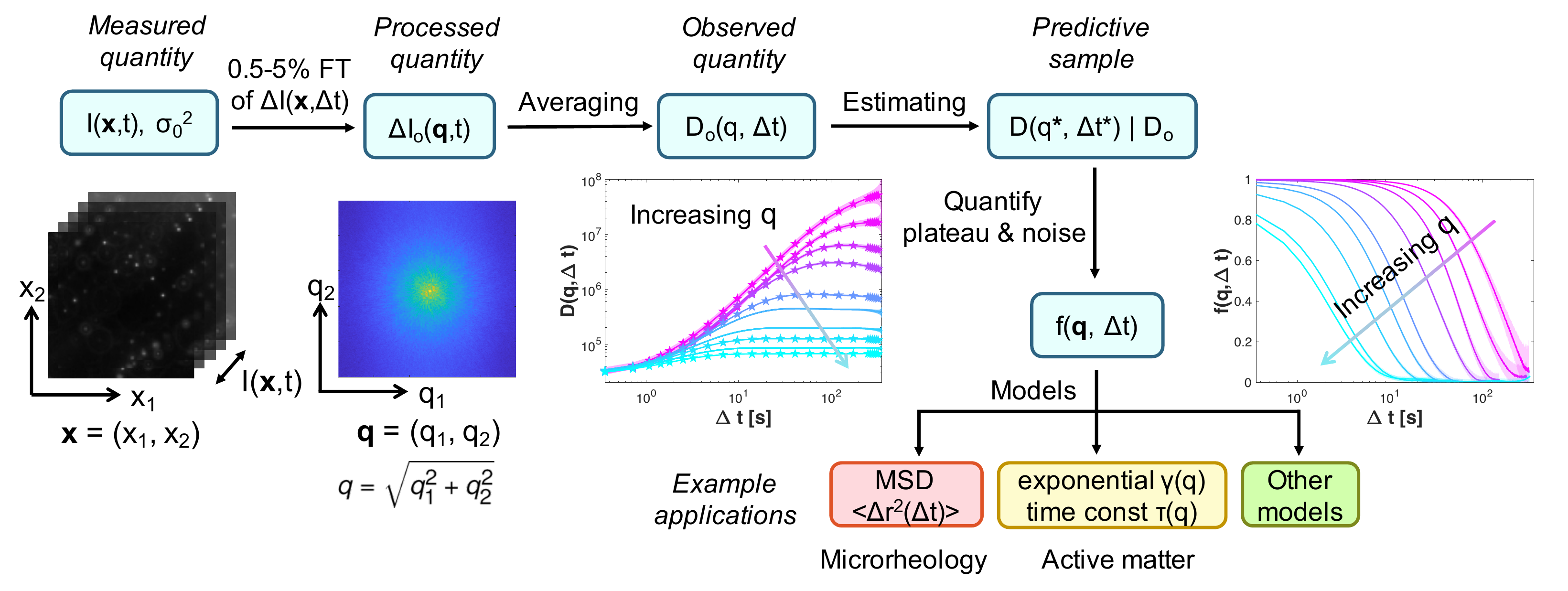}
    \caption{ Schematic representation of DDM-UQ-based data reduction, sampling, and fitting procedure used to determine material constants. From the stack of images acquired by microscopy, around $0.5\%$--$5\%$ of the Fourier transforms are performed to obtain the observed image structure function $D_o(q,\Delta t)$, from which a predictive sample is generated to estimate  $D(q^*,\Delta t^*)$ at unobserved $q^*$ and $\Delta t^*$, given $D_o(q,\Delta t)$. In the graphs of $D_o(q,\Delta t)$ versus $\Delta t$ and $f(q,\Delta t)$ versus $\Delta t$, at select $q$'s, the asterisks denote data selected for fitting, the lines denote values at all $\Delta t$'s for a particular $q$. The shadow denotes 95\% predictive interval, which is small compared to the range of the change in $D(q,\Delta t)$ over the range of $\Delta t$. The predictive samples preserve the quantiles of the distribution after transformation and are then used to find material quantities of interest.}
    \label{fig:flowdiagram}
\end{figure*}

In this work, we make three contributions towards overcoming these challenges of DDM: (1)
We relate the mean and variance of the error in the observed image structure function to the variance of the error in the original image intensities. By propagating the error, we show that there exists potential bias in different estimators of the noise and amplitude parameter in DDM, and we propose a new way to detect and reduce such bias. 
(2) We speed up the computation by using Gaussian process regression (GPR)  \cite{rasmussen2006gaussian} to overcome the computational bottleneck introduced by the Fourier transformations by subsampling the data at selected $\Delta t$ (Figure \ref{fig:flowdiagram}).  (3) Finally, we use the median of the predictive samples from GPR to robustly estimate the ISF, MSD and other quantities of interest. Furthermore, 
we illustrate through a broad range of examples, both simulated and experimental, how the choice of the estimators impacts  the accuracy of the resultant MSD and other quantities of interests.  We demonstrate that accurate estimation of the noise in image intensities is critical for  obtaining an accurate estimation of dynamical information in  DDM.              We make available a user-friendly MATLAB software package \cite{DDMUQmatlab} that implements fast computational techniques with uncertainty quantification developed in this work.

\section{Methodology}

We name our  algorithm \textit{differential dynamic microscopy with uncertainty quantification} (DDM-UQ). The analysis routine is described schematically in Figure \ref{fig:flowdiagram}. First, image stacks $I({\bf x},t)$ are acquired with a microscope or are produced using a particle dynamics simulation algorithm. The variance of the background noise intensity, $\sigma_0^2$, is either assessed independently or estimated from the image stack. Then, a small subsample of  a few percent  of the image differences are squared and Fourier transformed to construct a set of observed quantities $D_o({\bf q},\Delta t)$. Thereafter, GPR is fit to  $D_o({\bf q},\Delta t)$ to obtain a predictive distribution $D(q, \Delta t)$. Eventually, the predictive samples for each $q$ are used to obtain predictive samples for quantities of interest, such as $f(q, \Delta t)$ and $\langle  \Delta r^2(\Delta t) \rangle$.  Through analysis of a number of simulated and experimental data sets, we demonstrate that our  approach not only reduces the computational time, but in many cases also improves the accuracy and robustness of estimation, as compared to previous DDM approaches and MPT. 

\subsection{Error quantification}
\label{subsec:UQ}

To develop a statistical approach to error quantification and analysis, we write the observed intensity as 
\begin{equation}
I_o(\mathbf x,  t)=I(\mathbf x, t)+\epsilon(\mathbf x, t),
\label{equ:I_o}
\end{equation}
where $I(\mathbf x,  t)$ is an (unknown) deterministic function of the observed sample and imaging system; $\epsilon(\mathbf x, t)$ is an independent random noise with mean zero and variance $\sigma^2_0$; the subscript ``$o$" denotes the observed value. Several artifacts are known to impact the accuracy of DDM and are expected to contribute to $\epsilon(\mathbf x, t)$, including camera detection noise, edge effects arising from the finite field of view \cite{giavazzi2017image}, and effects of finite exposure time \cite{kurzthaler2018probing}. Others are known to affect MPT, such as the depth of field \cite{savin2007statistical} and finite pixel size \cite{savin2005static}, and are expected to impact DDM as well. Here we consider $\epsilon(\mathbf x, t)$ to be the difference between the measured signal and the ``true" intensity $I(\mathbf x, t)$ at each pixel without regard to the actual physical origin of the error. We note that there are other known spatially or temporally correlated artifacts, such as illumination fluctuations, that do not satisfy the criteria assumed for $\epsilon(\mathbf x, t)$. These will not be considered in the present analysis.

We illustrate how the error in Eq. (\ref{equ:I_o}) propagates in the analysis of DDM. The derivation of Eqs. (\ref{equ:representation_squared_intensity_fourier})-(\ref{equ:V_D_q_delta_t}) is given in Appendix A. 
Assuming that Eq. (\ref{equ:I_o}) holds, we can express the observed squared intensity function in reciprocal space as 
\begin{align}
&|\Delta \hat I_o({\bf q},t,\Delta t)|^2\nonumber \\
&=|\Delta \hat I({\bf q},t,\Delta t)|^2 + 2\Delta \hat I({\bf q},t,\Delta t)\Delta \hat \epsilon(\mathbf q, t,\Delta t)\nonumber \\
&\quad +|\Delta \hat \epsilon(\mathbf q, t,\Delta t)|^2,
\label{equ:representation_squared_intensity_fourier}
\end{align}
where the closed form expressions of $|\Delta \hat I({\bf q},t,\Delta t)|^2$, $ \Delta \hat I({\bf q},t,\Delta t)\Delta \hat \epsilon(\mathbf q, t,\Delta t)$, and $|\Delta \hat \epsilon(\mathbf q, t,\Delta t)|^2$ are given in Eqs. (\ref{equ:I_q_fourier})-(\ref{equ:I_epsilon_fourier}) in Appendix A, respectively.  

The expected value (i.e., mean) of $|I_o({\bf q},t,\Delta t)|^2$ is given by
\begin{equation}
\E[|\hat I_o({\bf q},t,\Delta t)|^2]=|\Delta \hat I({\bf q},t,\Delta t)|^2+2\sigma^2_0.
\label{equ:I_o_fourier_squared_expection}
\end{equation}

Note that the mean of the cross-product term, $\langle\Delta \hat I({\bf q},t,\Delta t)\Delta \hat \epsilon(\mathbf q, t,\Delta t)\rangle$, is zero under the assumptions made for $\epsilon(\mathbf x, t)$. 

By combining Eqs. (\ref{eq: nqnt}) and  (\ref{equ:representation_squared_intensity_fourier}), we can express the observations of the dynamic image structure function as follows:
\begin{align}
D_o(q,\Delta t)&=\langle |\Delta \hat I(\mathbf q,t, \Delta t)|^2\rangle  + 2\langle\Delta \hat I(\mathbf q,t, \Delta t) \Delta \hat \epsilon( \mathbf q, t, \Delta t)\rangle\nonumber \\
&\quad +\langle|\Delta \hat \epsilon(\mathbf q, t, \Delta t)|^2\rangle,
\label{equ:D_o_q_delta_t}
\end{align}
where $\langle \cdot \rangle$ denotes the ensemble with respect to $\mathbf q \in \mathcal S_q$ and $t\in \mathcal S_{\Delta t}$. The mean  of $D_o(q,\Delta t)$ follows:
\begin{align}
 \E[D_o(q,\Delta t)]&=D(q,\Delta t)+2\sigma^2_0. \label{equ:E_D_o}
  \end{align}
Further assuming $\epsilon(\mathbf x, t) \sim \mathcal N(0, \, \sigma^2_0)$ independently, we can calculate the variance
  \begin{align}
  \V[D_o(q,\Delta t)]&=\frac{2\sigma^2_0}{n_q n_{\Delta t}}\bigg( 2\sigma^2_0+2D(q,\Delta t)  \nonumber \bigg.\\
 &  \hspace{-.6in} \left. +\mbox{max}\left(0,(T-2l)\right) \left(\frac{\sigma^2_0}{n_{\Delta t}}-  \frac{2S_{q,\Delta t}}{(T-2l )n_{\Delta t}n_q}\right) \right),
 \label{equ:V_D_q_delta_t}
\end{align}
where the expression of $S_{q,\Delta t}$ is given in Eq. (\ref{eq:S_q}) in Appendix A, and $l=\frac{\Delta t}{\Delta t_{min}}$ is a positive integer small than $T$, with $T$ being the number of images and $\Delta t_{min}$ being the time lag between two consecutive time points.

The result in Eq. (\ref{equ:V_D_q_delta_t}) is  intuitive: the ratio $\frac{1}{n_qn_{\Delta t}}$ arises from the fact that the $D_o(q,\Delta t)$ is averaged from $n_q$ and $n_{\Delta t}$ observations of $\langle \hat I({\bf q},t,\Delta t)\vert^2 \rangle$, as shown in Eq. (\ref{eq: nqnt}), which decreases the variance. The other terms arise from the covariance between the $\mbox{sin}$ and $\mbox{cos}$ terms from the Fourier transform and that of the recursive sampling of the same image in different $\Delta t$.

Note that by Eq. (\ref{equ:E_D_o}),  an unbiased estimator of $D(q,\Delta t)$ is $D_o(q,\Delta t)-2\sigma^2_0$, while  using the observations $D_o(q,\Delta t)$ alone typically overestimates the image structure function by $2\sigma^2_0$ on average. Potential practical procedures for estimating $2\sigma^2_0$ will be discussed later.

We have shown that $D_o(q,\Delta t)$ can be separated into a deterministic term of the signal, $D(q,\Delta t)$, and a random term containing the noise and cross product of the noise and signal. This representation can be related to Eq. (\ref{equ:D_o_q_t}) by letting 
\begin{equation}
B(q,\Delta t) =2\langle\Delta \hat I(\mathbf q,t, \Delta t) \Delta \hat \epsilon( \mathbf q, t, \Delta t)\rangle +\langle|\Delta \hat \epsilon(\mathbf q, t, \Delta t)|^2\rangle, 
\end{equation}

\noindent where the mean and variance 
\begin{align}
    \E[B(q,\Delta t)]&=2\sigma^2_0, \\
        \V[B(q, \Delta t)]&=\V[D_o(q,\Delta t)],
\end{align}
where $\V[D_o(q,\Delta t)]$  is given in Eq. (\ref{equ:V_D_q_delta_t}). We observe that specifying $B$ as $0$ typically underestimates the mean of the noise term.  On the other hand, specifying $B$ as the average of the high-$q$ limit of observed image structure function $\langle D_o(q_{max},\Delta t)\rangle_{\Delta t}$, or as the ensemble average of static power spectrum $ 2\langle\vert \hat{I}_o (q_{max}, t)\vert^2\rangle_t$ \cite{giavazzi2018tracking},  tends to overestimate the mean of the noise by an small amount  $\langle A(q_{max})[1-f(q_{max},\Delta t)]\rangle_{\Delta t}$. Note that $[-f(q_{max},\Delta t)]$ is  close to 1 for  large $\Delta t$. Thus the bias is non-negligible when   $A(q_{max})$ is large. Also note that $A(q_{max})$ typically increases when the number of objects in the image or their peak intensity increases, when the image pixel size increases, or when the object size decreases; the relationship of $A(q)$ to some of these quantities was considered in \cite{bayles2016dark}. Indeed, we found that for a system with a large number of small objects, the bias can be large (Fig. \ref{fig:EstB}, Appendix C). When the pixel size is large, such as in the case of the actively driven system considered in Sec.  \ref{subsec:active_actin}, one may also tend to overestimate the noise. Similarly, using $D_{min}(q,\Delta t_{min})$, i.e., the minimum value of intermediate scattering function at the smallest $\Delta t$,  may also overestimate  $B$  in these scenarios. We found that the overestimation by $B_{est}=D_{min}(q,\Delta t_{min})$ may be smaller than the ones by $B_{est}=\langle D_o(q_{max},\Delta t)\rangle_{\Delta t}$ or  $B_{est}=2\langle\vert \hat{I}_o (q_{max}, t)\vert^2\rangle_t$, as the term $(1-f(q,\Delta t_{min}))$ can be close to zero. Furthermore, both estimators may slightly underestimate the noise due to its stochastic nature. For instance, when the signal contained in $D_o(q,\Delta t_{min})$ is close to zero, the ensemble average across $q's$ serves as a good estimator, whereas the minimum tends to underestimate the noise.
 In these cases, $D_o(q,\Delta t)$ tends to underestimate the noise much less frequently than overestimating the noise.

\begin{table*}[t]
\caption{ Potential bias in estimation approaches of noise $B_{est}$ in DDM.}
\centering
%\begin{tabularx}{\textwidth}{lX}
\begin{tabular}{ c|c } 
 \toprule
 $B_{est}$ & Scenarios for non-negligible bias \\
 \midrule
$0$ & underestimation in all scenarios  \\
$\langle D_o(q_{max},\Delta t)\rangle_{\Delta t}$    & overestimation when particle number, peak intensity, or pixel size is large, or if particle size is small \\     $ 2\langle\vert \hat{I}_o (q_{max}, t)\vert^2\rangle_t$& overestimation in similar scenarios as the above  \\
$D_{min}(\Delta t_{min})$  & overestimation when $\left(1-f(q,\Delta t_{min})\right)$ is not close to zero in similar scenarios as the above
\\  

\bottomrule
\end{tabular}
\label{tab:potential_bias_B}
\begin{flushleft}
% \textit{N.D. = not determined}
\end{flushleft}
\end{table*}

We summarize the limitations for each of the available approaches in estimating the mean of the noise in DDM in Table \ref{tab:potential_bias_B} and offer a simple test to detect the bias in Appendix C. In such a scenario, one may change experimental conditions, such as by reducing the number of objects, or increasing the object size, reducing the pixel size, etc. to reduce the bias; alternatively, the noise can be measured independently, or a more accurate estimator of the noise may be used. One goal of this exercise is to illustrate the importance of noise quantification, and to provide a way to detect potential bias in a wide range of scenarios.

For applications of DDM to microrheology, and assuming dilute probes with diffusive particle dynamics involving Gaussian displacements, we can relate the MSD $\langle  \Delta r^2(\Delta t)\rangle$ at each $q$ to the image structure function and related quantities as follows \cite{bayles2017probe,nijboer1966time}:
\begin{align}
&f(q,\Delta t) \nonumber \\
&\quad = \exp\left(-\frac{q^2\langle  \Delta r^2(\Delta t)\rangle}{4}\right)\left[1+\frac{\alpha_2 q^4\langle  \Delta r^2(\Delta t)\rangle^2}{32}+\cdots\right],
\label{equ: fqt r2 truncation}
\end{align}
 where the first order non-Gaussian parameter $\alpha_2 = \frac{d \langle  r^4(\Delta t)\rangle}{(d+2)\langle   \Delta r^2(\Delta t)\rangle}-1$ ($d$=dimensionality) is a measure of the heterogeneity or non-diffusive dynamics of the sample \cite{weeks2000three}. As is common in microrheology, we assume the contribution of the non-Gaussian parameter is negligible \cite{mason1997particle}, and thus $  \Delta r^2(q, \Delta t)$ at each $q$ may be  estimated  from $D(q,\Delta t)$ through the following approach:
\begin{align}
  \Delta r^2_{est}(q, \Delta t) &= \frac{4}{q^2}\mbox{ln} \left[ \frac{A(q)}{A(q)-D_o(q,\Delta t) + B(q,\Delta t)} \right]. \nonumber \\
\label{equ:msd}
\end{align}
It is common to assume sample ergodicity, and use the ensemble average of $  \Delta r^2(q, \Delta t)$ to estimate the mean squared displacement $\langle  \Delta r^2( \Delta t)\rangle$.  However, we note that the variance of $D_o(q,\Delta t)$ given in Eq.  (\ref{equ:V_D_q_delta_t}) is not the same across different values of $q$ and $\Delta t$. A simple ensemble average over wave vectors without addressing the weights due to different variances could introduce substantial bias in the estimation, when $D(q,\Delta t)$ at different wave vectors is not the same. We found that using the median, instead of the mean, can be more robust in such estimations. Numerical comparison between these estimation approaches will be discussed in the context of the simulated and experimental studies.  

\begin{figure}[htb!]

\includegraphics[height=.36\textwidth]{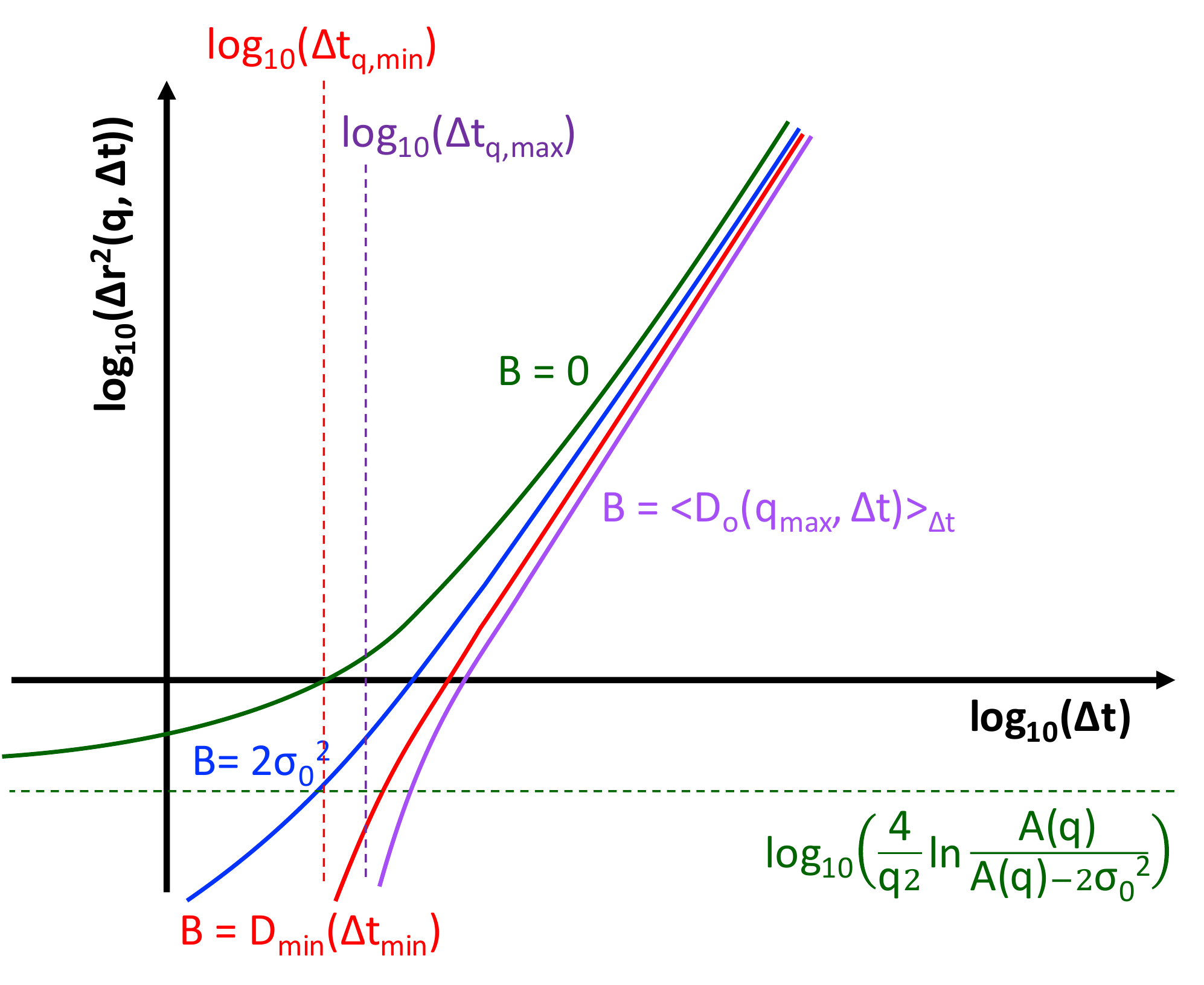}

	\caption{Illustration of how  four   different approaches to estimating $B(q)$ influence the calculated MSD. For given $q$, the quantity $\Delta t_{q,min}$ satisfies $D_o(q, \Delta t_{q,min})=D_{min}(\Delta t_{min})$.
	}
	\label{fig:comparison_sigma_2_0}
\end{figure}

To illustrate the importance of proper error estimation in estimating MSD, we consider the outcomes using four different ways of estimating $B$, graphed in Figure \ref{fig:comparison_sigma_2_0}.
If there exists $ \Delta t_{q,min}$, such that $D_o(q, \Delta t_{q,min})  = D_{min}( \Delta t_{min})$ for a given $q$,  and $B$ is estimated by $D_{min}(\Delta t_{min})$ (red solid line), then from Eq. (\ref{equ:msd}), the argument of the natural log approaches unity, and $\Delta r^2(q,\Delta t_{q,min}) \rightarrow 0$, which in turn drives  $\mbox{log}_{10}(  \Delta r^2(q,\Delta t_{q,min}))\rightarrow -\infty$, as shown.  {Choosing $B = \langle{D}_o (q_{max}, t)\rangle_t$ (purple solid line) will similarly overestimate the noise at small $\Delta t$, when $A(q)$ does not sufficiently approach 0 at $q_{max}$. }
% =D_{avg}(q_{max},\Delta t)

On the other hand,  if $B$ is estimated to be $0$ (green solid line), then approximately  $\mbox{log}_{10}(  \Delta r^2(q,\Delta t \rightarrow 0))\rightarrow \log_{10}\left[ \frac{4}{q^2}\mbox{ln} \left(\frac{ A(q) }{ A(q)-2\sigma^2_0}\right)\right]$  (an asymptotic value denoted by the green dotted line), as the expected value of $D_o(q,0)$, $\E (D_o(q,0)) = 2\sigma^2_0$. When we use the correct estimator $B = 2\sigma^2_0$ (blue solid line), the estimated $\Delta r^2(q,\Delta t) \rightarrow 0$ when $\Delta t\rightarrow 0$. 

%\end{comment}

These asymptotic limits demonstrate the crucial importance of properly estimating the mean of $B$ by  $\sigma^2_0$ when extracting dynamical properties from DDM such as the ISF or MSD, particularly at small $\Delta t$, a result that is further validated in  simulated and experimental examples (see e.g. Fig. \ref{fig:EstB} and Fig. \ref{fig:simulated_test_23}). 
A similar point about $B$ was noted previously by other researchers: while uncertainty in $A(q)$ is considered to dominate the analysis because it pertains to the signal, overestimating $B(q)$ such as in \cite{bayles2017probe} can lead to spurious results when computing $\langle \Delta r^2(\Delta t)\rangle$ \cite{edera2017differential}, and hence other authors proposed an iterative scheme to solve for $\langle \Delta r^2(\Delta t)\rangle$.  Treating $B$ as a fitting parameter was also commonly used in prior studies \cite{cerbino2008differential,wilson2011differential,lee2021myosin}, although the optimized value of $B$ can be  unstable in some scenarios, particularly if the weights of  $D_o(q,\Delta t)$ are not properly accounted for (see e.g. Fig \ref{fig:GL}). Through various simulation and real experiments in this study, we advocate that accurately estimating the noise parameter is crucial in DDM.   
In experiment, the noise term, $\sigma_0^2$, may be measured, typically through independent experiments using immobilized particles under identical imaging conditions, if a large bias of the estimator is detected. The variance of the image difference is then computed to give $\sigma_0^2$. In MPT, a ``noise floor'' is commonly quantified \cite{savin2005static} and frequently subtracted from all data to give a more realistic estimate of the MSD \cite{jawerth2020protein,furst2017microrheology,mcglynn2020multiple}.  
We show that while many estimators have negligible bias in approximating $2\sigma_0^2$ in many cases, other times it may be necessary to evoke an independently measured $2\sigma_0^2$, and we propose that noise characterization should be routine for DDM as well.

\subsection{ Gaussian Process Regression }
The  second challenge of DDM is the computational bottleneck that arises when performing a massive number of Fourier transformations.
We overcome this problem by representing the logarithm of image structure function by a Gaussian process regression (GPR) approach on a fraction ($0.5\%$--$5\%$) of data. GPR is a widely used machine learning tool for estimating nonlinear, smooth response surfaces and the predictions from GPR is the equivalence to  the kernel ridge regression (KRR) estimator, \cite{rupp2012fast,bartok2013representing,chmiela2017machine,brockherde2017bypassing,chmiela2018towards,bartok2018machine,wilkins2019accurate,anderson2019magma,wu2020emulating}.
%See e.g. \cite{wu2020emulating} for a brief review of  the connection between GPR and KRR, as well as their applications in physics. 

 We apply GPR to the logarithm of $D_o(q,\Delta)$, denoted  $\tilde D_o(\bm { \theta})=\ln(D_o(\bm {\theta}))$, with two input parameters: the natural logarithm of the wave vector (in reciprocal space) and time are denoted by $(\tilde q, \Delta  \tilde t)= (\ln(q),\ln(\Delta t))=\bm{ \theta}$. After obtaining the  predictive samples of the logarithm of the image structure function, we transform it back to obtain predictive samples for the image structure function. Because the logarithm of the ISF is smoother, the GPR approach works much better using logarithm of $D_o(q,\Delta t)$ as observations. Assuming that $n$ observations $\tilde D_o( {\bm \theta}_i)$, $i=1,2,...,n$ are used,  predictions by GPR can be represented through the following optimization, which simultaneously penalizes mean squared errors of the estimation with respect to the observations, as well as the complexity of the estimation: 
\begin{equation}
\hspace{-.11in}
 \tilde D_*=\underset{\tilde D\in \mathcal H}{\mbox{argmin}} \left\{\frac{1}{n}\sum^n_{i=1}[\tilde D_o( {\bm \theta}_i)-\tilde D( {\bm \theta}_i)]^2+\lambda || \tilde D ||^2_{\mathcal H} \right\}, 
\label{equ:f_krr}
\end{equation}
where   $||\cdot||_\mathcal H$ denotes the reproducing kernel Hilbert spaces regression (RKHS) norm (or native norm) \cite{rasmussen2006gaussian} that penalizes the complexity of the estimation to avoid overfitting and $\lambda$ is a regularization parameter. For any ${\bm \theta}_* $, the solution of (\ref{equ:f_krr}), known as KRR, is a weighted average of observations
\begin{align}
 \tilde D_*(\bm {\theta}_*)& = \mathbf w^T \mathbf {\tilde D}_o=\sum^n_{i=1} w_i\mathbf {\tilde D}_o(\bm \theta_i) ,
\label{equ:D_*}
\end{align}
where  $\mathbf w=(w_1,\dots,w_n)=\mathbf r^T_{\bm {\theta}_*} \mathbf{\tilde R}^{-1} $ is a row vector of weights, where $ \mathbf{\tilde R}=\mathbf R+n\lambda \mathbf I_n$ with $I_n$ being an identity matrix of size $n$ and $\mathbf R$ is an $n\times n$ correlation with $(i,j)$th entry parameterized by a kernel function $K(\bm \theta_i, \bm \theta_j)$, and $\mathbf r_{\bm \theta_*}=(K(\bm \theta_1,\bm \theta_*),...,K(\bm \theta_n,\bm \theta_*))^T$ is the correlation between predictive output and observations. 

Note that Eq. (\ref{equ:D_*}) is only a point estimator without giving assessment of  uncertainty. One advantage of the GPR approach  is the uncertainty of estimation can be quantified in a probabilistic framework. We model the latent function $\tilde D_o(\cdot)$  by a Gaussian process with noises, meaning that any marginal distribution $\mathbf {\tilde D}_o=(\tilde D_o(\bm \theta_1),...,\tilde D_o(\bm \theta_n))^T$ at $n$ inputs $\{ \bm \theta_1,...,\bm \theta_n\}$ follows a multivariate normal distribution:
\begin{equation}
\left((\tilde D(\bm \theta_1),...,\tilde D(\bm \theta_n))^T  \mid \mathbf m, \tilde {\mathbf R}, \sigma^2 \right) \sim \mathcal{MN} \left( \mathbf m,  \sigma^2  \tilde {\mathbf R} \right), 
\end{equation}
where $\mathbf m=(m(\bm \theta_1),..., m(\bm \theta_n))^T$ is a vector of the mean [assumed to be a constant in this work, i.e. $\mathbf m=(m,...,m)^T$] and $\sigma^2$ is a variance parameter. 

The power (stretched) exponential covariance function and Mat{\'e}rn covariance function  are often used for GPR \cite{rasmussen2006gaussian}. For any two inputs $\bm \theta_a=(\tilde q_a, \Delta  \tilde t_a)$ and $\bm \theta_b=(\tilde q_b, \Delta  \tilde  t_b)$,  we use a product covariance function $\sigma^2 K(\bm \theta_a, \bm \theta_b )=\sigma^2K_1(\tilde q_a, \tilde q_b)K_2(\Delta \tilde t_a, \Delta \tilde t_b)$, with $K_l(\cdot,\cdot)$, $l=1,2$, following a Mat{\'e}rn correlation with roughness parameter $5/2$ such that   
\begin{equation}
K_l(x_a,x_b )=\left(1+{\sqrt{5}\beta_l d}+ \frac{5\beta^2_l d^2 }{3} \right)\exp\left( -{\sqrt{5}\beta_l d }\right),
 \label{equ:matern_5_2}
\end{equation}
 where $d=| x_a- x_b |$ for any real valued input $x_a$ and $x_b$ with inverse range parameter $\beta_l \in \mathbb R^+$, $l=1,2$. The sample path of the Gaussian process with Mat{\'e}rn correlation in (\ref{equ:matern_5_2}) is twice  differentiable and is often used as a default correlation in GPR \cite{gu2018robustgasp}.  

The parameters in GPR $(m, \sigma, \bm \beta, \lambda)$ can be estimated by the maximum likelihood approach discussed in Appendix B. Plugging in the estimated parameters $(m_{est},  \sigma^2_{est},   {\bm \beta}_{est},  \lambda_{est} )$, the predictive distribution of $\tilde D(\bm \theta_*)$ at any $\bm \theta_*$ follows a normal distribution  \cite{rasmussen2006gaussian}:  
\begin{align}
(\tilde D(\bm \theta_*)\mid \tilde{\mathbf D}_o )\sim \mathcal N( \tilde D_{est}(\bm \theta_*),   \sigma^2_{est} K_*(\bm \theta_*, \bm \theta_*)+\tilde \sigma_*^2),
\label{equ:pred_dist}
\end{align}
where $\tilde \sigma_*^2=\sigma^2_{est}\lambda_{est}$ is the variance of the noise $\epsilon_{\bm \theta_*}$, and
\begin{align}
\tilde D_{est}(\bm \theta_*)&=  m_{est} + \mathbf r^T_{\bm \theta_*} \mathbf{\tilde R}^{-1} (\mathbf {\tilde D}_o-  m_{est}  \mathbf 1_n ) \label{equ:pred_median},\\
K_*(\bm \theta_*, \bm \theta_*)&=\sigma^2_{est} \left(K(\bm \theta_*, \bm \theta_*)-\mathbf r^T_{\bm \theta_*}
\mathbf{\tilde R}^{-1}\mathbf r_{\bm \theta_*}\right).
\label{equ:pred_variance}
\end{align}

We use the predictive median $\tilde D_{est}(\bm \theta_*)$  for predicting the logarithm of the ISF to obtain the unsampled $\bm \theta_*$ and the predictive median of $D_o(\bm \theta_*)$ can be obtained by transforming the predictive median of  $\tilde D_o(\bm \theta_*)$ through the  exponential function. 
The predictive median in Eq. (\ref{equ:pred_median}) is equivalent to KRR  in Eq.  (\ref{equ:D_*}) when the mean parameter is zero $m_{est}=0$. Here the uncertainty of predictions and predictive samples can be obtained by the predictive distribution in Eq. (\ref{equ:pred_dist}).

\begin{figure*}[t]
\includegraphics[width=0.9\textwidth]{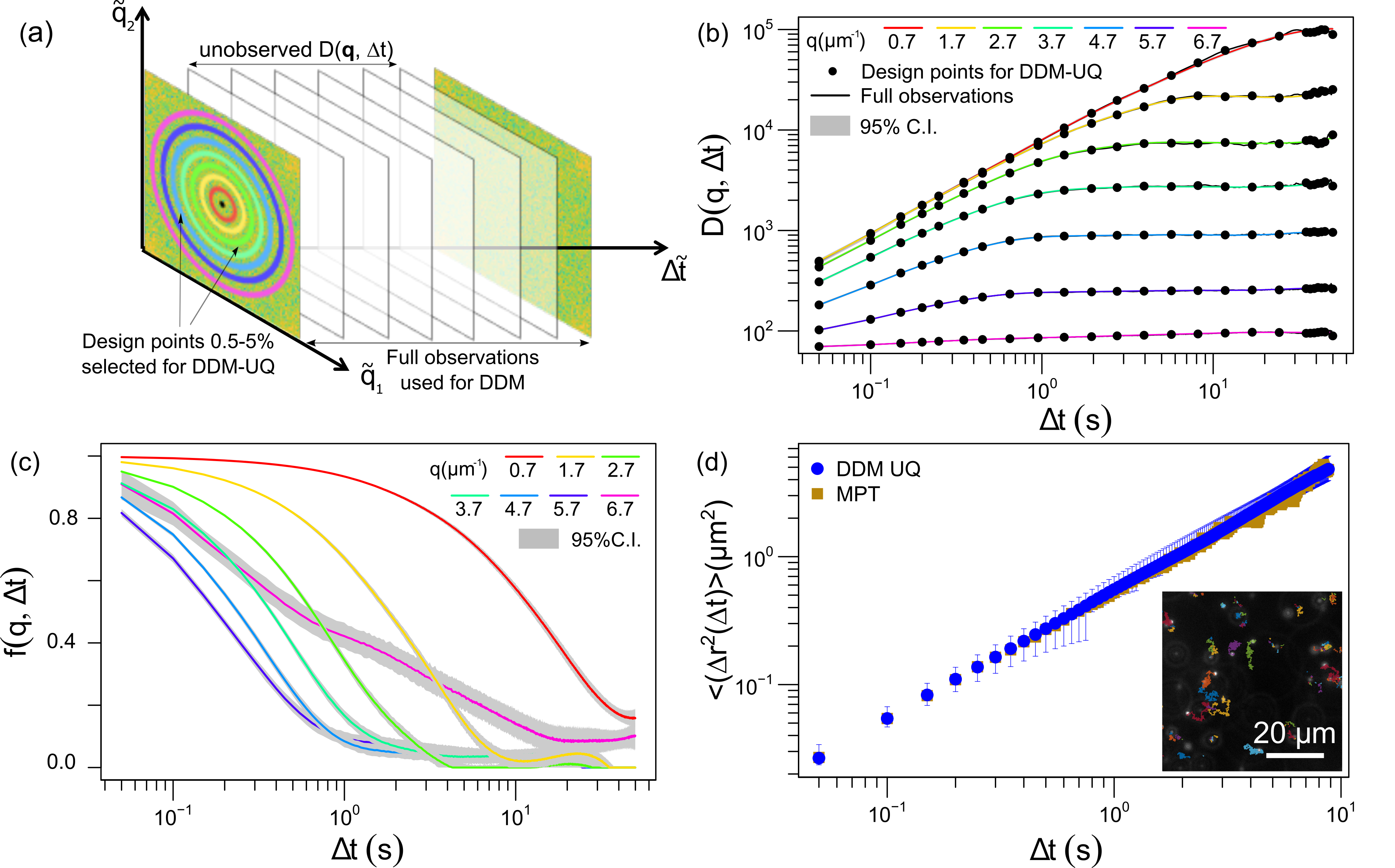}

	\caption{(a) The design points (blue circles) are selected along the logarithmically scaled coordinates $(\mathbf q, \Delta t)$, requiring only a fraction of the Fourier transformations to compute $D_o(q, \Delta t)$. (b) Predictive median and predictive samples by GPR for representative values of $q$'s. The variance of noise is proportional to the inverse of the number of pixels in ensemble. 300 predictive samples  for each $q$ were generated to obtain the 95\% credible interval (gray shadow) . The observations used for GPR are plotted as black dots. The full observations (black lines) overlap with the predictions (colored lines).(c) The predictive samples for the intermediate scattering function are plotted against the $\Delta t$. (d) Predictive mean squared displacement is plotted against the $\Delta t$ using all three methods. The $95\%$ predictive interval is shown with the error bar for DDM-UQ. The inset shows a snapshot of experiment with MPT trajectory overlay. Plots (b)--(d) are derived from a movie with 1 $\mu$m probe particles diffusing in a viscous fluid.}
	\label{fig:comparison_samples}
\end{figure*}

\subsection{Predictive sampling by a  downsampled data set}

After obtaining the image intensities, DDM-UQ analysis starts by selecting a fraction of the $\Delta t$'s equally spaced logarithmically along the input space coordinates to compute $D_o(q,\Delta t)$ 
as shown in Fig. \ref{fig:comparison_samples}(a).  Here, we choose not to downsample observations at wave vectors ($q$), but the approach can be extended to $q$ as well, if one wishes to further reduce the computational cost.   The Fourier transformation is only performed on this reduced set of image differences, allowing for fast, high-throughput analysis. Since an estimate of the plateau of the image structure function at long times is often required by the analysis, we ensure that at least five design points lie on the interval $[0.7\Delta t_{max}, 0.9 \Delta t_{max}]$ on the $\Delta \tilde t$-axis for each chosen $\tilde q$ [Fig. \ref{fig:comparison_samples}b].  For all numerical results analyzed in this study, we only use $D_o(q,\Delta t)$  at 25 $\Delta t$ points in DDM-UQ, which represents around $0.5\%$ to $5\%$ of total observations.   
The predictive median of $D_o(q,\Delta t)$ is smoother than the observed  $D_o(q,\Delta t)$ since spurious noise is effectively filtered out [see e.g. the black curves and colored curves in Fig \ref{fig:comparison_samples}b].

Note that our goal is to relate $D_o(q, \Delta t)$ to other quantities of interest, such as the intermediate scattering function $f(q,\Delta t)$ or mean squared displacement $\langle   \Delta r^2(\Delta t)\rangle$, both of which are nonlinear transformations of $D_o(q, \Delta t)$. To achieve this, we sample image structure functions from the predictive distribution in Eq. (\ref{equ:pred_dist}) and transform the samples to obtain other quantities of interest. The transformed predictive samples can be used to estimate the predictive interval of quantities of interest at a given set of input $(q, \Delta t)$.  In this work, we sample $1000$ observations, denoted as $D_s(q,\Delta t)$ for $s=1,2,\dots,1000$ at any $(q,\Delta t)$, and transform $D_s(q,\Delta t)$ to obtain other quantities of interest, such as intermediate scattering function $f_s(q,\Delta t)$ via Eq. (\ref{equ:D_o_q_t}) after estimating $B_{est}$ and $A_{est}(q)$.  The determination of these two estimators is discussed in Sec.  \ref{subsec:numerical_setting}.   For each $(q,\Delta t)$, the $95\%$ predictive interval of $f(q,\Delta t)$ can be estimated by the lower $2.5\%$ quantile and the upper $2.5\%$ quantile of the transformed predictive samples $f_s(q,\Delta t)$ for $s=1,2,\dots,300$. 

Since the predictive samples of $D(q,\Delta t)$ preserve information such as quantiles of distribution for any transformation, transforming these samples can be used to  estimate the ISF $f(q,\Delta t)$ and MSD $\langle   \Delta r^2(\Delta t)\rangle$  as shown in Figs. \ref{fig:comparison_samples}(c) and \ref{fig:comparison_samples}(d).   The predictive median is used for estimating MSD at each $\Delta t$ as it is typically more robust than the mean.  Conditional on the observed values $D_o(q,\Delta t)$, the predictive samples of $f(q,\Delta t)$ and  $\langle   \Delta r^2(\Delta t)\rangle$  can be used to  assess estimation uncertainty as well. 

\section{Validation through simulation}

\subsection{Image formation and analysis approach}
\label{subsec:numerical_setting}
To validate our methodology, we first employ simulations of a time series of images demonstrating particle motion. This approach has the significant advantage that the true particle motion is known \textit{a priori} and the true MSD has a closed form expression, thus allowing quantitative comparison with results obtained using DDM, DDM-UQ, and MPT, the latter obtained using an open source tracking algorithm \cite{gao2009accurate, crocker1996methods}. Moreover, it is possible to investigate systematically how different sources of noise influence algorithm performance, and whether different functional forms of $\langle   \Delta r^2(\Delta t)\rangle$ perform differently.

We first examine simulated video images of particles in motion, each with a Gaussian intensity profile with peak intensity $I_c$ = 255 and standard deviation of $\sigma_p$ = 2 pixels. In principle, the intensity recorded at a single pixel $I_p({\bf x})$ could arise from intensity contributions of multiple particles in the vicinity. The contribution of the $j$-th particle located in ${\bf x}_j(t)$ is given by:
\begin{equation}
    I_p({\bf x},{\bf x}_{j}(t)) = I_c exp\left(-\frac{({\bf x}-{\bf x}_j(t))^2}{2\sigma_p^2}\right).
\end{equation}

To account for noise in the background intensity signal, a time-varying, random uniform noise $I_b({\bf x},t)$, centered around zero, in the range $[-10, 10]$ is added. Thus, we may compute $\sigma_0^2$ as the variance of the background noise $I_b({\bf x},t)$: $\sigma_0^2 = \frac{20^2}{12}\approx 33.3$. Thus, the signal at a time $t$ is the sum of signals attributed to all particles as well as the background:
\begin{equation}
    I({\bf x},t) = I_b({\bf x},t)+\sum_{j=1}^{n_p} I_p({\bf x}-{\bf x}_j(t)).
    \label{signal}
\end{equation}

Note that the pixel intensities in simulation are not subjected to a cut-off ceiling value as are those obtained in imaging (e.g. 0-255 for an 8-bit image). Moreover, unlike MPT, where the relative brightness of particle and background significantly affects tracking precision \cite{savin2005static}, in DDM signal quality depends sensitively on the magnitude of the image difference. Keeping this context in mind, we simulate particles with brightness $I_p({\bf x}-{\bf x}_j(t))$ that does not vary with time, and we vary the step size by which the particles move in each time step instead. It is well-known that DDM performance deteriorates, and can completely break down, in the limit of small probe displacements \cite{bayles2017probe}. For diffusive particles taking a step with a variance $\sigma_s^2$, we compare the performance of DDM, DDM-UQ, and MPT in calculating the MSD values. 

The DDM-UQ analysis represents data obtained using our proposed approach based on the downsampled Fourier transformation of $\Delta I({\bf x},\Delta t)$'s with GPR and predictive samples.   For all numerical results that extract the MSD using DDM-UQ, we estimate $B_{est}$ to be the minimum of $D_{min}(\Delta t_{min})$ and $\langle D_o(q_{max},\Delta t)\rangle_{\Delta t}$, to reduce the overestimation bias, as summarized in Table  \ref{tab:potential_bias_B}. The estimate of the amplitude parameter $A_{est}(q)$ is obtained by $A_{est}(q)+B_{est}=2\langle I^2_o(q,t)\rangle_{t}$, as it applies to all wave vectors, regardless of whether or not a plateau is reached at the given $\Delta t$. Furthermore, extracting the MSD from the ISF at extremal wave vectors could lead to large errors. This is because the  intensities of only a few pixels are used to calculate the ISF at small wave vectors, leading to large uncertainty in estimation. On the other hand, $A(q)$ can approach zero at large wave vectors, leading to unstable estimate of MSD in this regime. To avoid these extremes, DDM-UQ uses $D_o(q,\Delta t)$ based on 80 intermediate $q$ values, typically ranging from the 4$^{th}$ up to 83$^{rd}$ largest $q$'s. 

Additionally, we record the MSD estimations for $\Delta t$ values that are no larger than $90\%$ of the estimated plateau values according to Eq. (\ref{equ:msd}). In this limit, the denominator is sufficiently different from zero to ensure a small variance of the estimation. Finally, we truncate the estimation of MSD at those  $\Delta t$  where fewer than 10 wave vectors are available, to avoid selection bias in estimation when sampling is limited.    Other approaches to truncating the wave vector range and lag time based on standard deviation of the data were explored in prior DDM analyses \cite{bayles2017probe}.  Weighting  $D(q,\Delta t)$ based on the variance of the data without truncation of wave vectors and time points could be an efficient way to reduce selection bias while allowing robust estimation of the extracted MSD at longer $\Delta t$. This is a potential future topic for research.

For DDM analysis in practice, we perform the Fourier transformation of all values of $\Delta I({\bf x},\Delta t)$, and estimate $B(q)$ using the  average of $D(q_{max},\Delta t)$ over all $\Delta t$'s at $q_{max}$, $\langle D(q_{max},\Delta t)\rangle$. Our first simulated case contains 800 small particles (Fig. \ref{fig:simulated_test_23}), whereas we include 50 moderately large particles in the latter 6 simulations. As shown in Figure \ref{fig:EstB}, all estimators of the mean of the noise $B$ for the six latter simulated examples are very similar, as expected. We then test how the DDM analysis is affected by  estimating $A(q)$ two different ways.  Following the procedures (I) from \cite{bayles2017probe} we estimate $A(q)$ from the plateau or (II) using the static power spectrum (i.e. $A_{est}(q)+B_{est}=2\langle I^2_o(q,t)\rangle_t$) per  \cite{escobedo2018microliter,cerbino2017dark}.  We largely follow the procedures defined by \cite{bayles2017probe, bayles2016dark} to select the wave vectors and $\Delta t$ values for estimation and data selection for both DDM analysis approaches.   

Also included is the MPT approach, which was performed by locating particles in each frame, and searching in the vicinity to link trajectories of individual particles. The localization error is not characterized, given the high-particle intensity compared to the background intensity in all cases investigated.  Note that for certain cases, such as optically dense cases of simple diffusive processes compared in Sec. \ref{subsec:validation_noise}, MPT cannot identify the large number of probes, so MPT results are not compared in these two studies.

In detail, in all cases, we generated videos of the motion of $n_p$ = 50 particles, except for the case described in Sec. \ref{subsec:validation_noise}, where $n_p$ = 800. The simulation box is a 2D square with sides $L$ = 480. The movie spans time steps $t=1,2,\dots,1000$. We further imposed displacements in each successive time step $\Delta x_{i,j}(t) = x_{i,j}(t+1)-x_{i,j}(t)$ (where $i$ = 1,2 stands for the $x_1$, or $x_2$ directions in the Cartesian coordinates in 2D). 

We construct three scenarios that represent the general features observed in a broad range of experiments: simple diffusion, diffusion with drift, and constrained diffusion within a harmonic potential well (i.e., an Ornstein–Uhlenbeck process). These scenarios will result in distinct shapes of $\langle   \Delta r^2(\Delta t)\rangle$ and highlight various challenges to each analytical approach. The derivations of the expected values of the MSDs for all simulated scenarios are given in Appendix D. 

As a metric of the accuracy of a given analysis method, we compare the normalized root mean squared error (N-RMSE) for the estimated MSD relative to the known true MSD: 
\begin{equation}
\mbox{N-RMSE}=\frac{\sqrt{\frac{1}{n_{\Delta t}}\sum_{\Delta t \in \Delta \mathcal T}[\langle\Delta \tilde{r}^2(\Delta t)\rangle- \langle   \Delta \tilde{r}^2_{est}(\Delta t)\rangle ]^2 }}{\tilde{\sigma}_{r}},
\end{equation}
where $\langle \Delta \tilde r^2(\Delta t)\rangle$ is the logarithm of the true MSD with  base 10, $\langle \Delta \tilde r^2_{est}(\Delta t)\rangle$ is the corresponding estimate using DDM  with two different $A(q)$ estimators,  DDM-UQ or MPT, and $\tilde \sigma_{r}$ is the sample standard deviation of the logarithm of the true MSD with  base 10. In practice, not all $\Delta t$ values are available for every method,  due to large fluctuations at large $\Delta t$ values, and this provides a limit to the total $\Delta t$ range captured in each case. To ensure that the four methods are evaluated on the same test set $\Delta \mathcal T$ and to ease quantitative comparisons, we determine the usable range of $\Delta t$'s by the smallest maximum $\Delta t$ available among the three methods. The N-RMSE of different simulated cases is summarized in Table \ref{table: RMSE for simulations}.

\subsection{Validating the use of $\sigma_0^2$ as an estimator for $B$}
\label{subsec:validation_noise}

We have shown that the mean of $B$ is $2\sigma^2_0$, and that the estimation of $\sigma^2_0$ is critically important to the analysis of DDM data. To illustrate this point, we first show the results of a simulation of $n_p=800$ particles with $\sigma_p$ =0.5 moving in a purely viscous fluid. We generate movies demonstrating simple Brownian motion: at each time step, $\Delta x_{i,j}(t)\sim \mathcal N(0,\sigma^2_s)$ independently, where $\sigma_s$ represents the step size with units of pixels. The expected MSD of the Brownian motion is $\E(\langle   \Delta r^2(\Delta t)\rangle)= 2\sigma^2_s \Delta t$. For 2D diffusive motion, we thus expect $\langle r^2(\Delta t)\rangle = 4D_m\Delta t$, where $D_m$ is the diffusion coefficient. $D_m$ can be associated with the step size $D_m = \frac{\sigma_s^2}{2}$. For $\sigma_s^2 = 4$, and $D_m = 2$, the truth is $\langle r^2(\Delta t)\rangle = 4D_m\Delta t = 8 \Delta t$.

This simulation represents an optically dense case with a large number of small particles, which can induce a large bias in estimating the noise (Table \ref{tab:potential_bias_B}). We demonstrate this by comparing the estimation of noise by  $B =D_{min}(\Delta t_{min})$,  $B =\langle D_o(q_{max},\Delta t)\rangle_{\Delta t} $ and $B= 2\langle\vert \hat{I}_o (q_{max}, t)\vert^2\rangle_t$ to the truth $B =2\sigma^2_0$ (horizontal line) as shown in Fig. \ref{fig:EstB}. We found that all three estimators overestimate the mean of $B$ in the analysis. The first approach, $B =D_{min}(\Delta t_{min})$, has the smallest bias among the three, as the term $1-f(q,\Delta t_{min})$ in Eq. (\ref{equ:D_o_q_t}) is relatively small at the smallest $\Delta t$. 

\begin{figure}
    \centering
    \includegraphics[width=.45\textwidth]{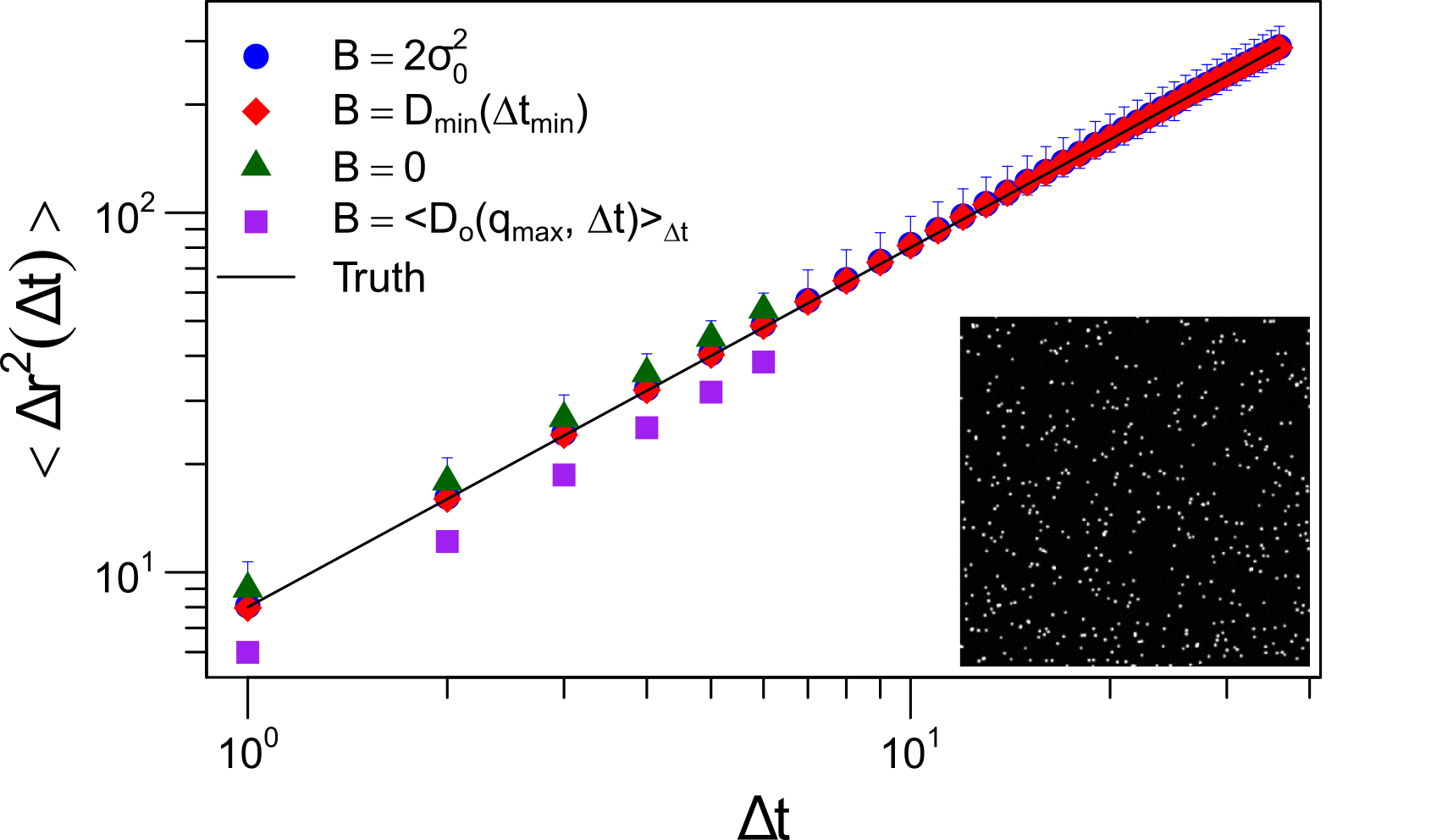}
    \caption{The ensemble-averaged MSD calculated from the simulated 2D Brownian motion of 800 particles.  
	Estimation of MSD by DDM-UQ with four different ways  of estimating the mean of the noise.  
	The error bars denote $95\%$ predictive interval of DDM-UQ with $B=2\sigma^2_0$. The truth of $\langle \Delta r^2(\Delta t)\rangle=8\Delta t$ is denoted as the black line. The inset shows the initial position of particles, where particles are enlarged for better visualization. 
	Note that it is not possible to perform MPT due to the large number of particles moving within the frame.}
    \label{fig:simulated_test_23}
\end{figure}

Next, we calculated the MSDs by DDM-UQ using each of the following ways for estimating the noise, as shown in Figure \ref{fig:simulated_test_23}:
 $B =D_{min}(\Delta t_{min})$ (red diamonds),  $B =\langle D_o(q_{max},\Delta t)\rangle_{\Delta t} $ (purple squares), $B =2\sigma^2_0$ (blue circles) and $B=0$ (green triangles). The truth is plotted as the thick black line. Our results reveal that estimating the noise accurately is necessary to obtain an accurate estimate of MSD, whereas overestimating or underestimating the noise leads to an inaccurate estimate, particularly for sufficiently small $\Delta t$, where the displacements are smaller and therefore more strongly impacted by a poorly estimated noise term.

A simple way to detect the estimation bias of $B$  is to plot $D_o(q,\Delta t)$ as a function of $q$ for a given $\Delta t$ (here selected to be $\Delta t_{min}$) across all $q$ values, as shown Fig. \ref{fig:DetectBias}. If the signal $A(q)(1-f(q,\Delta t))$ does not approach zero at the highest $q$ values, then $D_o(q,\Delta t)$ continues to decrease as $q$ increases. We found this to be the case for $D_o(q,\Delta t)$ for the optically dense case (with $N_p=800$ probes, shown as filled circles, Fig. \ref{fig:DetectBias}). Hence, estimating $B$ using $D_o$ at $q_{max}$ introduces non-negligible bias in this example. In contrast, when there are only $N_p=50$ particles (shown as open circles), the $D_o(q,\Delta t)$ approaches zero at high-$q$ and the bias for $B$ is is negligible across all methods of estimation (Fig. \ref{fig:EstB}).  
To overcome the bias induced by the estimator, one may adjust the experimental conditions to avoid the scenarios summarized in Table  \ref{tab:potential_bias_B}, or may attempt to measure the noise using a separate sample with probes immobilized in solid matrix under similar imaging conditions. Deriving a more accurate estimator of the noise at these experimentally challenging scenarios will be  an interesting future direction. 

\subsection{Simple diffusion with different step size $\sigma_s$}

\begin{figure*}[t]
\includegraphics[width=0.9\textwidth]{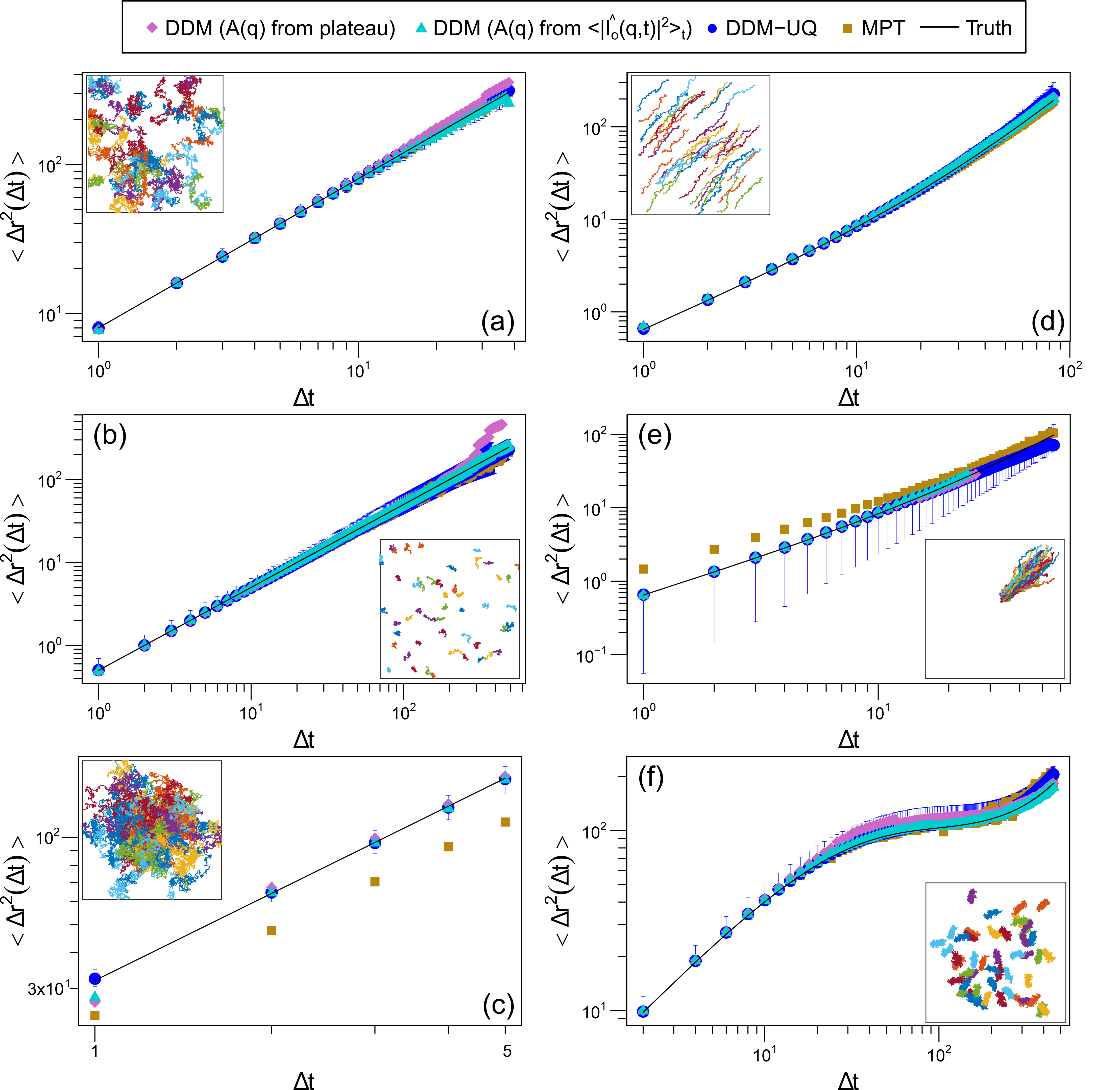} 
	\caption{Comparisons of MSDs calculated using DDM-UQ, DDM, and MPT for various simulated scenarios: (a)--(c) Simple diffusion with (a) $\sigma_s=2$, (b) $\sigma_s=0.5$ and (c) $\sigma_s=4$, corresponding to intermediate, low,  and high step sizes. (d-e) Diffusion with drift in (d) optically dilute and (e) optically dense samples, and (f) diffusion of a particle within a harmonic potential well  (Ornstein-Uhlenbeck process) subjected to drift.  The pink diamonds and the cyan triangles indicate mean values obtained by estimating $A(q)$ from the plateau in $D(q,\Delta t)$, or from $\langle\vert \hat{I}_o (q, t)\vert^2\rangle_t$, respectively, using DDM algorithm based on all values of $D(q,\Delta t)$.  The blue circles and error bars depict the mean and $95\%$ predictive interval estimated by DDM-UQ based on $1\%$ of the $D(q,\Delta t)$, respectively. The golden squares and black solid line represent the output of MPT analysis and the true values as directly calculated from the known particle positions, respectively. The generated particle trajectories are shown in the insets, in a simulation box of $480\times480$ pixels.}
	\label{fig:comparison_test_simulation}
\end{figure*}

To explore the effects of varying step size (which is a proxy for varying diffusivity) on our analysis, three additional scenarios with diffusive dynamics are explored, using simulations of particles taking different step sizes $\sigma_s$, but for which all other settings and conditions were held constant. When $\sigma_s=2$, corresponding to an intermediate step size, all four methods (DDM with two different estimators, DDM-UQ and MPT) provide results that reasonably approximate the true values that are directly calculated from the inputted particle positions (Fig. \ref{fig:comparison_test_simulation}a). When the N-RMSE values are calculated and compared, we find that the results from MPT provide the best approximation of the true values for this case, whereas DDM-UQ provide nearly the same level of accuracy as MPT (Table \ref{table: RMSE for simulations}).

\begin{table*}[t]
\caption{N-RMSE of simulated cases. The analysis mode with the lowest N-RMSE is shown in bold in each scenario. }

\centering
\begin{tabular}{ l|c|c|c|c } 
 \toprule
 Scenario & DDM ($A(q)$ from plateau) & DDM ($A(q)$ from $\langle\vert \hat{I}_o (q, t)\vert^2\rangle_t$) & DDM-UQ & MPT \\
 \midrule
Simple diffusion, $\sigma_s=2$ & 0.109 & 0.047 & 0.025 & \bf{0.024} \\
Simple diffusion, $\sigma_s=0.5$  & 0.396 & {\bf 0.034} & 0.057 & 0.129 \\
Simple diffusion, $\sigma_s=4$  & 0.130 & 0.095 & \bf{0.014} & 0.492 \\  
Diffusion with drift & 0.061 & {0.052} & 0.074 & {\bf 0.026} \\
Diffusion with drift, optically dense &0.041 & 0.049& \bf{0.029} & 0.424\\ 
O-U process with drift & 0.243 & \bf{0.063} & 0.183 & 0.213 \\  

\bottomrule
\end{tabular}
\label{table: RMSE for simulations}
\begin{flushleft}
% \textit{N.D. = not determined}
\end{flushleft}
\end{table*}

We next calculated the MSDs for  simple diffusion with lower ($\sigma_s=0.5$) and higher ($\sigma_s=4$) step sizes [Figs. \ref{fig:comparison_test_simulation}(b) and \ref{fig:comparison_test_simulation}(c)].
The largest differences are observed at high step sizes, [Fig. \ref{fig:comparison_test_simulation}(c)] where particle displacements are large ($\sigma_s=4$). In this limit, DDM-UQ  outperforms  MPT by  a large margin (Table \ref{table: RMSE for simulations}).  The reason is intuitive for MPT: as particle displacement becomes large, the likelihood of two or more particles  exchanging positions within the search radius increases significantly. This can lead to the algorithm misidentifying the particle, thereby resulting in erroneous linking of the trajectories.

\subsection{Diffusion with drift}

We next consider particles subjected to diffusive-convective motion (i.e. ``drift''). At each time step, a particle moves by $\Delta x_{i,j}\sim \mathcal N(\mu_D,\sigma_s^2)$, resulting in a random walk with step size $\sigma_s = 0.56$ superimposed on a deterministic drift with mean velocity $\mu_D = 0.1$ in units of pixels/time step in the same direction for all particles. The expected value of the MSD in this case is $\mathbb E[\langle   \Delta r^2(\Delta t)\rangle]=2\sigma^2_s \Delta t+2\mu^2_D\Delta t^2$. At short $\Delta t$, the motion is primarily diffusive [Fig.  \ref{fig:comparison_test_simulation}(d)], $\langle   \Delta r^2(\Delta t)\rangle \sim \Delta t$  with a transition to convective motion, $\langle   \Delta r^2(\Delta t)\rangle \sim \Delta t^2$, at large $\Delta t$. This results in an  increasing value of $\frac{d\langle   \Delta r^2(\Delta t)\rangle}{d\Delta t}$ with $\Delta t$, and permits a measure of the relative strength of the two dynamic processes through measure of the local slope on a log-log scale. 

We also compare the performance of the different analytical routines under differing initial conditions.  In particular, we vary the initial positions where particles were released at $t = 1$, holding all other settings equal, to compare an optically dilute scenario [Fig. \ref{fig:comparison_test_simulation}(d)], where particles were uniformly distributed throughout the simulation box, and an optically dense scenario [Fig. \ref{fig:comparison_test_simulation}(e)], where particles were released from a small $\frac{L}{20} \times \frac{L}{20}$ square in the middle of the frame causing them to be near each other and even overlap for some frames. When particles are evenly distributed in the simulation box, all methods closely track the truth, with MPT having the lowest RMSE  in the optically sparse scenario, whereas DDM-UQ performs the best in the optically dense scenarios. The optically dense scenario mimics situations where a high concentration of particles is present, which is known to lead to tracking issues in MPT since it is often difficult to distinguish particles in close proximity.  

By contrast, DDM and DDM-UQ perform strongly in this limit (Table \ref{table: RMSE for simulations}). The range of $\Delta t$ that can be resolved by any method decreases as compared to the dilute case, due to the large variability of MSD at different wave vectors in this scenario. 

\subsection{Ornstein–Uhlenbeck process with drift}
Finally, we simulated particles from an Ornstein–Uhlenbeck (O-U) process with drift. Such a process mimics thermally-driven particle motion in an effective elastic medium with drifts distinct to each particle.  The convective term is constant in magnitude $\mu_D$, and fixed in direction $\theta$ for an individual particle, but randomized for all particles. A pure O-U process without a convective term can lead to a $\langle  \Delta r^2(\Delta t)\rangle$ that plateaus at a certain $\Delta t$, and thus $f(q,\Delta t)$ does not decorrelate for some finite values of $\Delta t$. Adding a convective term leads to complete decorrelation and allows the dynamics at these $\Delta t$'s to be captured.
Moving the sample to achieve ensemble averaging for sample that manifested constrained heterogeneity has previously been applied to light scattering on polymer gel samples \cite{xue1992nonergodicity}. 

Here, just like the original O-U process, successive steps have a weak correlation with previous steps:
\begin{equation}
    X_{i,j}(t+1)  =  \rho X_{i,j}(t) +\epsilon_{i,j,t},
    %+ \mathrm{step}
\end{equation}
where $\epsilon_{i,j,t}\sim \mathcal N(0,\sigma^2_s(1-\rho^2))$, with $\rho=0.95$,  $\sigma_s=5$ and $X_{i,j}(t) = x_{i,j}(t) -(t-1)\mu_{ij} -x_{i,j}(1)$ for any $t$; the indices represent the $j$-th particle and $i$-th direction ($i$ = 1, 2) and $\mu_{1j} = \mu_D \cdot cos\theta_j$, $\mu_{2j} = \mu_D \cdot sin \theta_j$, respectively, with  $\mu_D=0.02$. Each particle's initial position was generated by a normal distribution centered around $x_{ij}(t_0)$ with variance $\sigma^2_s$, i.e. $x_{ij}(t_1) \sim \mathcal N(x_{ij}(t_0), \sigma^2_s)$, where $x_{ij}(t_0)$ are 
randomly distributed within a square $\frac{3L}{4} \times \frac{3L}{4}$ in the middle of the simulation box to reduce the likelihood that a  particle moves out of the frame during the simulation. The motion is subject to an attractive potential towards $x_{ij}(t_0)+(t-1)\mu_{ij}$ with drift  $(\mu_Dcos(\theta_j), \mu_Dsin(\theta_j))^T$ at the $t$th time point for the $j$th particle. The expected value of MSD for O-U process with drift is $\mathbb E[\langle   \Delta r^2(\Delta t)\rangle]=4\sigma^2_s(1-\rho^{\Delta t}) +\mu^2_D\Delta t^2$.  There is a diffusive contribution that dominates at sufficiently small $\Delta t$ from the first term, $1-\rho^{\Delta t}\approx 1-(1+ln(\rho)\Delta t)= ln(\rho^{-1})\Delta t$, by Taylor expansion at small $\Delta t$. The second convective term captures the convective flow, and can represent the type of dynamics observed in some actively driven systems \cite{monnier2012bayesian,lee2021myosin}. 

The process results in a trace of MSD versus $\Delta t$ with multiple inflection points: $\langle   \Delta r^2(\Delta t)\rangle$ first grows with decreasing slope until the  plateau value is reached ($\frac{d\langle   \Delta r^2(\Delta t)\rangle}{d\Delta t} \approx 0)$; at much longer $\Delta t$'s, the MSD will grow with an increasing slope until eventually the convective motion dominates ($\frac{d\langle   \Delta r^2(\Delta t)\rangle}{d\Delta t}$ = 2). In contrast to all previous scenarios, we find that $D(q,\Delta t)$ exhibits non-monotonic behavior  at some wave vectors, and that the MSD shows apparent $q$-dependence over $\Delta t$ for many intermediate $q$ values. Specifically, a local maximum in $D(q,\Delta t)$ appears before $D(q,\Delta t)$ reaches the plateau at long $\Delta t$. Such non-monotonic behavior in the resulting $f(q,\Delta t)$ has previously been observed by in a system consisting of self-catalytic Janus particles \cite{kurzthaler2018probing}.

We find that all approaches are able to capture the general trends and magnitude of motion [Fig. \ref{fig:comparison_test_simulation}(f)]. At short $\Delta t$, the response of all four methods is similar. We note that this scenario constitutes an ``ideal'' case for MPT: the particle locations are sparse, and each particle moves toward their respective attractive centers, $x_{ij}(t_0)$, never to cross paths with one another. Yet DDM-UQ still outperforms MPT in this case, even though it can be shown that some marginal benefit can be gained by using the full data set. Still, with limited sampling, DDM-UQ has performance on par with DDM and MPT, evaluated by the N-RMSE  in Table \ref{table: RMSE for simulations}.

Overall, we find that DDM-UQ accurately determines $\langle  \Delta r^2(\Delta t)\rangle$ for a range of experimentally-relevant scenarios based on a fraction of the observations,  which substantially reduces the computational cost. Note that DDM-based algorithms are more automated compared to MPT, as DDM does not require manually chosen inputs of model parameters such as particle sizes and search radius. On the other hand, MPT can provide  more reliable dynamic information at larger $\Delta t$ than the current model-free DDM-based algorithms. The DDM-UQ algorithm developed in this study enables a model-free automated DDM-based analysis with results that are comparable to MPT with optimized settings, with less computational cost and tuning parameters. Indeed, DDM-UQ even outperforms MPT for certain challenging experimental scenarios (e.g. high concentration, fast moving objects, etc). Furthermore, many other scenarios abound where the ISF $f(q,\Delta t)$ rather than the MSD can provide physical insight to the system, and the DDM-UQ algorithm provides an automated, model-free estimation of the ISF.  These findings affirm the sensitivity of this ensemble-based method as well as the need for an unbiased estimator of noise and other model parameters, and validates our data reduction approach.

\section{Analysis of experimental data}

\subsection{Newtonian fluid}
\label{subsec:Newton_fluid}
We first measured the properties of a Newtonian fluid in which we expect simple diffusive particle dynamics. Experimentally, we suspended fluorescent polystyrene microspheres of diameter 2$a$ = 100 nm (yellow-green with excitation maxima of $\lambda_{ex}$ = 441 nm and emission maxima at $\lambda_{em}$ = 485 nm, Polysciences, Warrington, PA) in a 30 wt\% sucrose solution (Sigma-Aldrich, St. Louis, MO) at a particle volume fraction $\phi \approx  3\times 10 ^{-5}$. This composition was previously studied by dark-field DDM \cite{bayles2016dark}, and the exact viscosity value of the sucrose is well documented \cite{swindells1958viscosities}.
The particle suspension is introduced into a home-made sample cell formed using a glass slide and glass cover slip separated by 100 $\mu$m spacers. The sample is imaged in epifluorescence using an Olympus IX73 inverted microscope, outfitted with a halogen lamp with green fluorescent protein (GFP) filter set ($\lambda_{ex}$ =  457--487 nm, $\lambda_{em}$ =  502--538 nm), using a 40$\times$ objective (NA = 0.6), which provides a spatial resolution of 97 nm/pixel. Images are collected using a 8-bit Point Grey Chameleon USB camera using a 100 ms exposure time, 10 Hz frame rate, and 960 pixel $\times$ 960 pixel frame size. 

\begin{figure}[h]
		
\includegraphics[width=.45\textwidth]{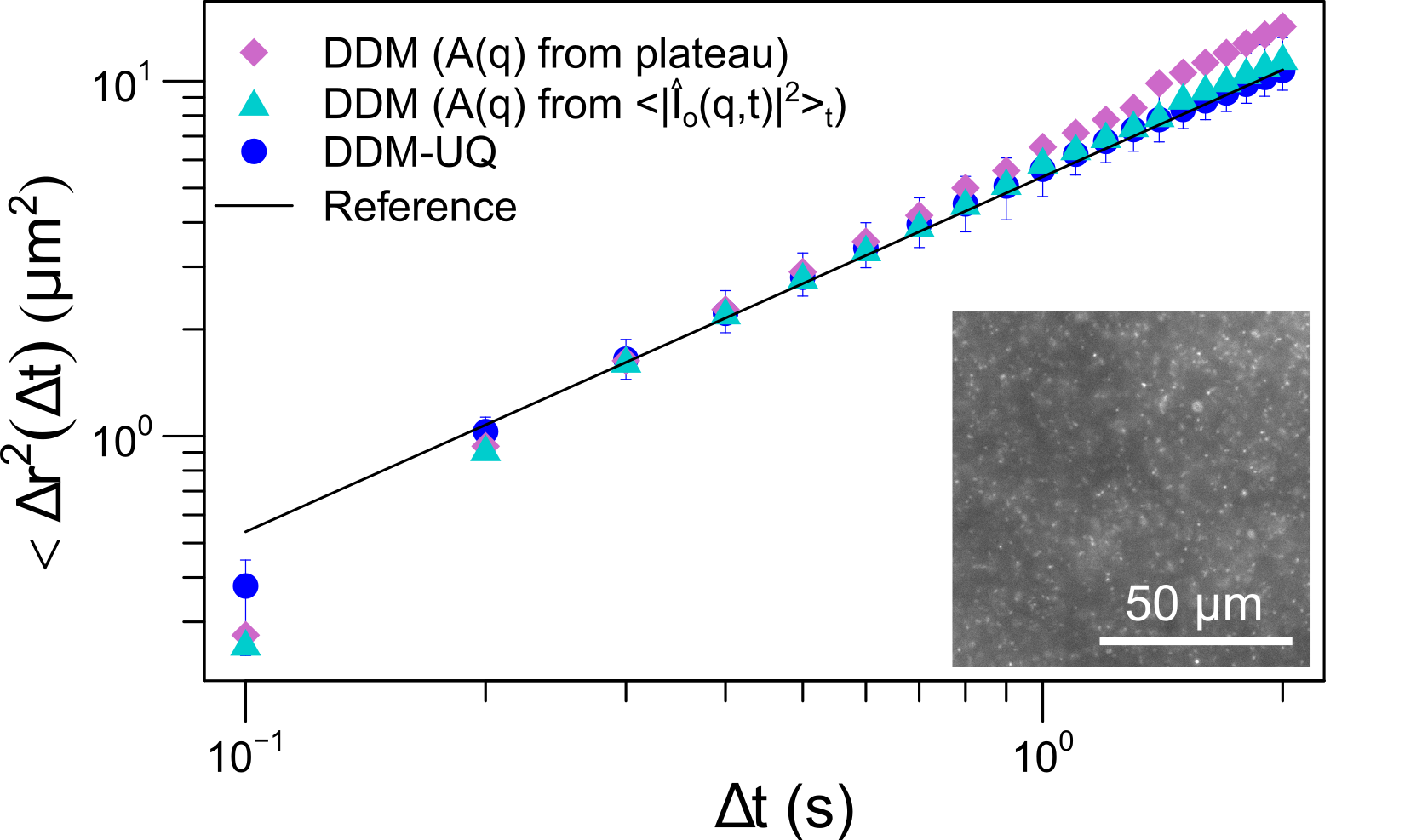} 
\caption{Mean squared displacements estimated from the motions of 100 nm diameter probe particles in a 30wt\% sucrose solution, which serves as a model Newtonian fluid.  The pink diamonds and the cyan triangles indicate MSD obtained by using $A(q)$ determined from the plateau in $D_o(q,\Delta t)$, or from $\langle\vert \hat{I}_o (q, t)\vert^2\rangle_t$, respectively, based on all values of $D(q,\Delta t)$.  Blue circles denote results obtained using the DDM-UQ analysis by a small fraction of the data; in this case $A(q)$ was only estimated using $\langle\vert \hat{I}_o (q, t)\vert^2\rangle_t$. The black solid line denotes reference values and is calculated using the known viscosity (at 20$^o$C) and the Stokes Einstein Equation. The inset shows a single experimental image of the movie, where particles appear to be grainy and cannot be individually resolved.}
	\label{fig:comparison_sucrose}
\end{figure}

In this experiment, the limited resolution of fluorescence microscopy relative to the particle size precludes identification of individual particles by MPT [see Fig. \ref{fig:comparison_sucrose}(a), inset], and thus prevents MPT analysis. Despite the lack of particle-level information, DDM is nevertheless capable of detecting the minute differences in image intensities due to particle motion, and recovers the correct diffusive dynamics [Fig. \ref{fig:comparison_sucrose}(a)]. Thus, DDM shows extraordinary sensitivity even when the particle size is below the diffraction limit of the microscope. 

The reference value of the MSD determined by the Stokes-Einstein relation \cite{bayles2016dark} is reported by the solid black line. We generally find quantitative agreement by both DDM and DDM-UQ over most of the measured $\Delta t$'s, with noticeable deviations from the linear trend and the expected values at the smallest $\Delta t$ (Fig. \ref{fig:comparison_sucrose}).    Here the effects of using the different estimators of noise are very small, as $ D_o(q, \Delta t)$ levels off at high $q$ values for any given $\Delta t$. 
When the approaches are compared, DDM-UQ (blue diamonds), which stabilizes large fluctuations through the use of the predictive median,
% uses an independently measured estimate of $\sigma_0^2$, better approximates the noise,
leads to more accurate estimates of the MSD. 
  Furthermore, we found that estimating $A(q)$ from  $A_{est}(q)+B_{est}=\langle\vert \hat{I}_o (q, t)\vert^2\rangle_t$  (blue circles and cyan triangles) improves the estimate of the MSD at large $\Delta t$ over cases where $A$ is estimated from the plateau values only (pink diamonds). We attribute this improvement to averaging more $D_o(q,\Delta t)$'s from  more $q$'s in determining the ISF, regardless of whether there is a plateau or not at the finite $\Delta t$ that are accessible in experiment.  

%\end{comment}

\subsection{Viscoelastic fluid}

\begin{figure*}
    \centering
    \includegraphics[width=0.98\textwidth ]{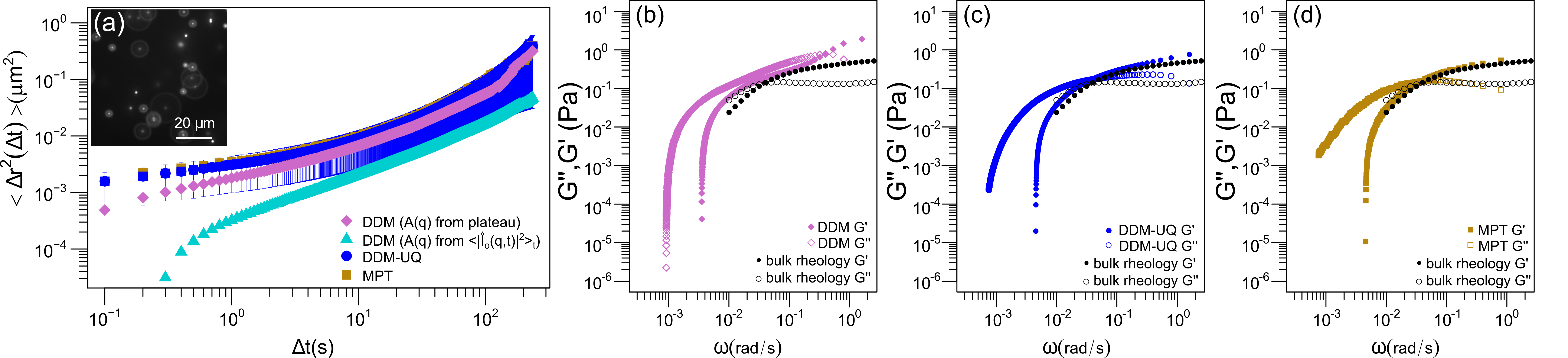}
    \caption{ Results of microrheology and bulk rheology measurements of solutions worm-like micelles, which form viscoelastic fluids. (a) Mean squared displacements obtained from  full $D_o(q,\Delta t)$ with $A(q)$ estimated from the plateau  in $D(q,\Delta t)$ (pink diamonds), from $\langle\vert \hat{I}_o (q, t)\vert^2\rangle_t$ (cyan triangles),  DDM-UQ (blue circles) and MPT (golden squares). The inset shows an experimental snapshot of the microrheology experiment. (b)--(d) Comparison of the frequency-dependent linear viscoelastic moduli obtained either from bulk rheology experiments (black symbols) or calculated from the MSDs obtained by either using (b) DDM with $A(q)$ estimated from $\langle\vert \hat{I}_o (q_{max}, t)\vert^2\rangle$ (pink diamonds), (c) DDM-UQ (blue circles), or (d) MPT (golden squares). Solid symbols denote the storage [elastic, $G^{\prime}(\omega)$] modulus while open symbols denote the loss [viscous, $G^{\prime \prime} (\omega)$] modulus.  MSDs estimated by MPT and by DDM-UQ nearly overlap. }
    \label{fig:GSER_WLM}
\end{figure*}

We next investigate the performance of DDM in probing the dynamics of a non-Newtonian fluid; namely, a viscoelastic worm-like micelle solution of 12.5 mM sodium salicylate (NaSal; Sigma-Aldrich, St. Louis, MO) and 15 mM cetylpyridinium chloride (CPyCl; Sigma-Aldrich, St. Louis, MO) that forms an entangled network. To this solution, fluorescent polystyrene microspheres of diameter 2$a$ = 1500 nm (carboxylated yellow-green ex/em = 505/515, Life Technologies, Carlsbad, CA) are added at a volume fraction $\phi \approx 2 \times 10^{-4}$. The sample is mixed and allowed to relax overnight prior to loading into a capillary tube, which is sealed on both sides with optical glue (Norland Products, Inc.) and cured under a UV lamp. We note that this sample is similar in composition, but not identical, to a solution characterized by DDM microrheology in previous work \cite{bayles2017probe}. The sample is imaged using a Zeiss Axio Observer 7 microscope in fluorescence mode using a Colibri 7 light source, standard GFP filter sets and a 40$\times$ water-immersion objective lens (NA = 1.2), which provides a magnification of 150 nm/pixel. Images were recorded with an Axiocam 702 monochromatic camera using  15 ms exposure time, 10 Hz frame rate, and 512 pixels $\times$ 512 pixels frame size. In this case, the reference data set is obtained by a bulk rheology measurement of an identical sample without tracer particles using an AR-G2 stress-controlled rheometer (TA Instruments, New Castle, DE) to perform a frequency-sweep in the linear viscoelastic limit using a 40-mm diameter cone-and plate fixture, with a 2$^o$ cone angle and a 55 $\mu$m truncation, at 2\% shear strain over a frequency range of 0.01--10 rad/s. The instrument is outfitted with a solvent trap to minimize evaporation during testing. 

Wormlike micelles (WLMs) manifest complex frequency-dependent viscoelasticity that nonetheless follow simple scaling laws \cite{rehage1991viscoelastic, bayles2017probe}. Such a system is challenging to characterize: 
small probe displacements at low $\Delta t$ make it difficult to determine if the flattening of the MSD at low $\Delta t$ is characteristic of system behavior or a result of ``pixel biasing'' due to particle localization error \cite{savin2005static}. The slow dynamics also contribute to the sensitivity of $q$-selection. In this case, solid-like behavior at high-frequency (low $\Delta t$) is confirmed by bulk rheometry measurements [Figs. \ref{fig:GSER_WLM}(b)--\ref{fig:GSER_WLM}(d)] and thus a ``flattening'' of the MSD trace is expected at small $\Delta t$. With this information, we evaluate the four methods.

In this experiment, the noise estimators $B_{est} = \langle D_o(q_{max},\Delta t)\rangle_{\Delta t}$  and  $B_{est} = D_{min}(\Delta t_{min})$  produce similar results by DDM, although neither method performs well because the estimation error for $A(q)$ at high values of wave vector is relatively large due to the small displacements, which approach the resolution limit at small lag times.  

\begin{figure}
    \centering
    \includegraphics[width=0.4\textwidth]{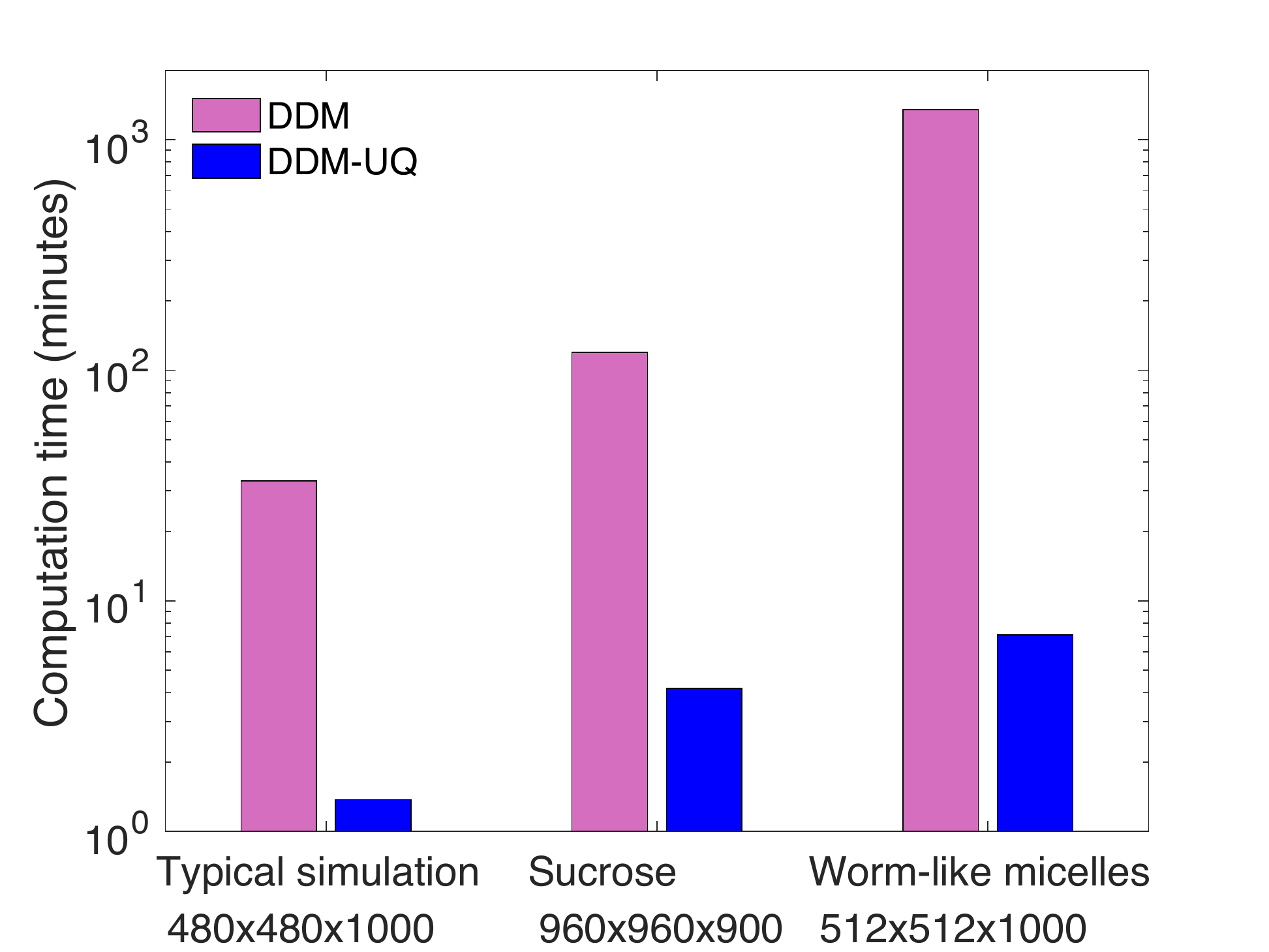}
    \caption{Pair bar graphs demonstrating the computation efficiency of DDM-UQ scheme (blue) over current DDM approaches (pink). The size of each image stack is labeled underneath the data set, in terms of frame size $\times$ time points ($L_x\times L_y\times T$).}
    \label{fig:TimeComp}
\end{figure}

By contrast, the DDM-UQ algorithm using the median from the predictive sampling based on $D_o(q,\Delta t)$ on moderately large wave vectors is more robust than using a simple ensemble of  $D_o(q,\Delta t)$ in DDM.   
The MSD trace for DDM-UQ is very close to the MPT result, and is more consistent with the Maxwell fluid-like behavior (Fig. \ref{fig:GSER_WLM}a).  Note that we only compute the Fourier transformation of intensity difference at 25 selected $\Delta t$'s in DDM-UQ analysis, instead of 6000 $\Delta t$ points in DDM, which reduces the computational cost by more than 100 times (shown in Figure \ref{fig:TimeComp}) while providing more accurate results.  

Next we convert the measured MSD data into measures of the frequency-dependent viscoelastic moduli using the generalized stokes Einstein relation (GSER) \cite{mason2000estimating}:
\begin{equation}
    \vert G^*(\omega)\vert \approx \frac{k_B T}{\pi a \langle \Delta r^2(1/\omega)\rangle\Gamma[1+\alpha(\omega)] },
\end{equation}

\noindent where $T$ is the temperature, $k_B$ is the Boltzmann constant, and $\alpha(\omega)$ the power-law slope [on a log-log plot of $\langle \Delta r^2(\Delta t)\rangle$]. 

The procedure of determining the power law slope $\alpha(\omega)$ usually involves first taking the numerical derivative of $\langle \Delta r^2(\Delta t)\rangle$ with respect to $\Delta t$, and then   fitting a polynomial of the data around a particular $\Delta t$, which is then Laplace transformed to frequency space \cite{mason2000estimating}. As shown in Fig. \ref{fig:GSER_WLM}(b)--\ref{fig:GSER_WLM}(d), the moduli we measured directly by bulk rheology generally agree with those computed from $\langle \Delta r^2(\Delta t)\rangle$ using microrheology approaches, even at high frequency, in both magnitude and in the frequency of the crossover. The result from DDM with $A(q)$ estimated from the plateau is shown in pink diamonds [Fig. \ref{fig:GSER_WLM}(b)], the result using DDM-UQ is shown in blue circles [Fig. \ref{fig:GSER_WLM}(c)], and MPT is shown in golden squares [Fig. \ref{fig:GSER_WLM}(d)].  The MSD by DDM with $A(q)$ estimated from the relation $A_{est}(q)+B_{est}=\langle\vert \hat{I}_o (q, t)\vert^2\rangle_t$ (cyan triangles) contains very large error,  and thus it fails to estimate moduli, so the result is not shown here.  

Both the frequency dependence and magnitude of the moduli in the high-frequency regime are extremely sensitive to the MSD at low $\Delta t$. Nevertheless, MPT and DDM-UQ are in an approximate agreement with the values obtained with macroscale rheology at higher frequencies.   By contrast, in a slow moving system, or at large $q$'s (which correspond to small displacements), when the movement is less than a pixel on average over $\Delta t$, it is not possible to capture the system dynamics by calculating the image difference. As a result, estimates of the MSD from the $D_o(q,\Delta t)$ tend to provide underestimates as compared to the true values at small $\Delta t$. 

We note that the numerical differentiation of $\langle \Delta r^2(\Delta t)\rangle$ introduces a high degree of uncertainty into the moduli that is not represented on Fig. \ref{fig:GSER_WLM}.  
A more robust estimation method for the moduli is an area of future research.

\subsection{Model fitting for actively-driven systems}
\label{subsec:active_actin}
Beyond thermally-driven dynamics and microrheology, DDM can be applied in the realm of non-Brownian, active systems where models of $f(q,\Delta t)$ can provide insight into the physics associated with the system dynamics.  
In such cases, the time-variation of the image structure function does not arise from probe motion but from structural evolution of actively-powered components within the system such as migrating bacteria \cite{martinez2012differential} or advection due to internal stresses that arise from phase separation and aging \cite{gao2015microdynamics}, among other examples. Here we demonstrate that DDM-UQ analysis can be applied to such anomalous dynamics as well. Specifically, the statistical approach we have developed should be applicable to any system that sufficiently decorrelates over the experimental observation time, thus greatly  reducing computational time and increasing  the robustness of model fits.

\begin{figure}
\includegraphics[width=0.35\textwidth]{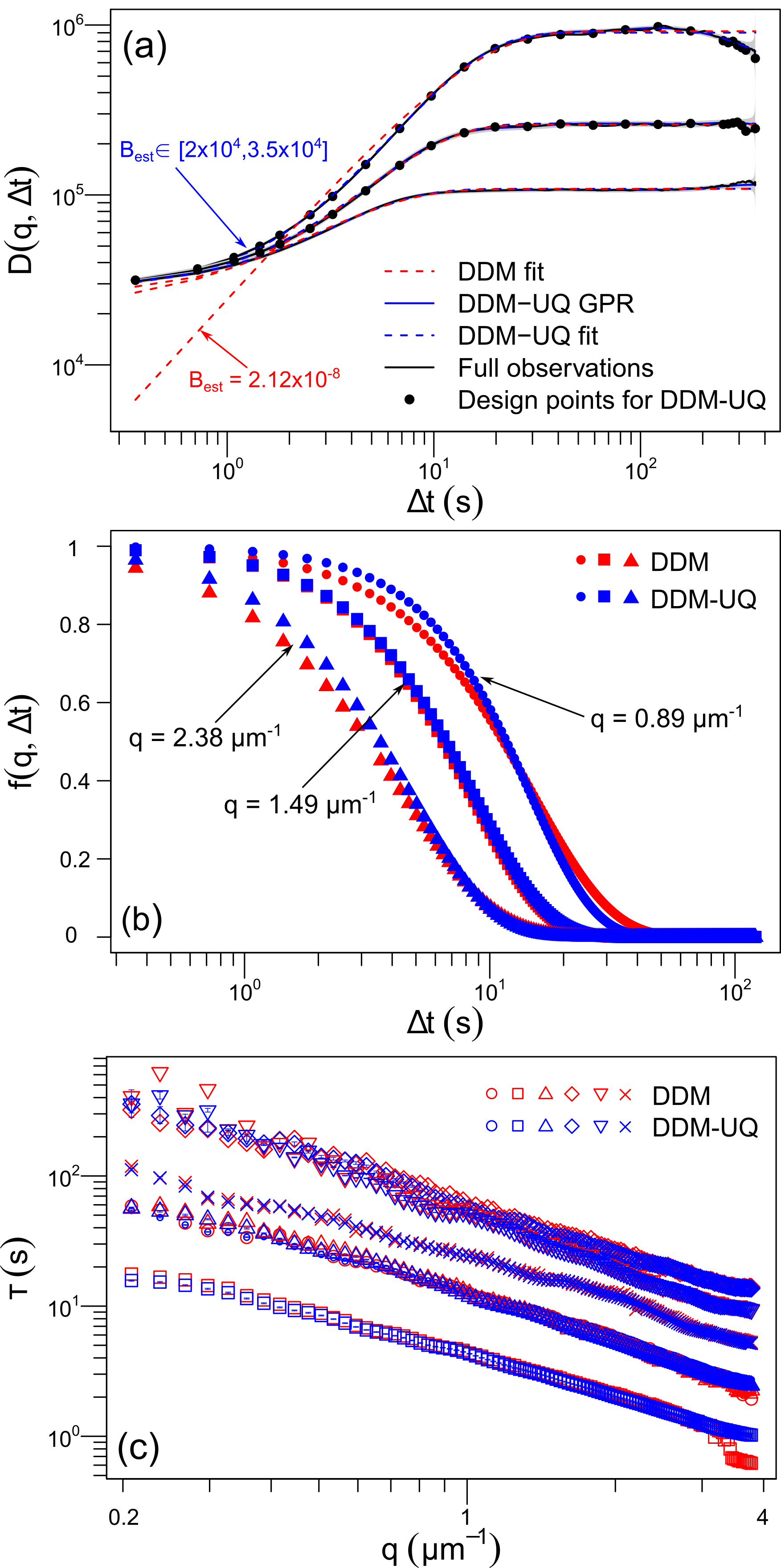}
\caption{Example of analysis of an actively driven system. Experimental data extracted from a fluorescently-labeled actin-myosin-microtubule composite network system reported in \cite{lee2021myosin}, processed using DDM (red symbols) or DDM-UQ (blue symbols), and fit to a stretched exponential model to describe the system dynamics [Eq. (\ref{eq:stretched exponential})]. (a) Comparing DDM, and DDM DQ fits to the original $D_o(q,\Delta t)$ matrix, where the uncertainty is denoted by the gray shaded area, which is very small as the uncertainty is low. The solid line denotes full $D_o(q,\Delta t)$ data, while the solid dots denotes $D_o(q,\Delta t)$ selected to obtain the predictive distribution.(b) Fits to $f(q,\Delta t)$ shown at several different $q$'s.  (c) Parameter $\tau$, which describes the system relaxation time, plotted as a function of $q$. Different symbols represent $\tau$ extracted from six different data sets. From this, the system velocity is obtained (Fig. \ref{fig: GL velocity MLE CI}).}
\label{fig:GL}
\end{figure}

As an example, we consider the dynamics of actively-driven composite cytoskeletal networks of actin and microtubules, by (re)analyzing the experimental data recently reported by Lee $et$ $al$. \cite{lee2021myosin}. Actin and microtubules are ubiquitous and essential in eukaryotic cells, and \textit{in vitro} networks of the purified filamentous proteins are widely studied for their potential to self-organize and form model non-equilibrium materials when acted upon by ATP-driven molecular motors such as myosin. In recent work, Lee $et$ $al$. \cite{lee2021myosin} investigated a composite network comprising actin and microtubules, which were each labeled with distinct fluorophores, and acted upon by myosin. DDM was then used as a means of disentangling the motions recorded through the individual fluorescence channels, allowing investigation of the mechanisms of cross-correlation of actin and microtubule dynamics within the entangled network. By contrast, MPT can provide only information on the bulk network itself. 

 Here we follow  \cite{lee2021myosin} to model the  dynamics of such an active system  by a stretched exponential model \cite{lee2021myosin}:
\begin{equation}
     f(q, \Delta t) = \exp\left(-\left(\frac{\Delta t}{\tau(q)}\right)^{\gamma (q)} \right),
     \label{eq:stretched exponential}
\end{equation}

\noindent where $\gamma (q)>1$ means the system shows contractile dynamics, while $\gamma (q)<1$ when the system shows stretching dynamics. A relaxation time scaling where $\tau(q)= \frac{1}{vq}$ describes a system exhibiting ballistic motion with velocity $v$, whereas $\tau(q)= \frac{1}{D_m q^2}$ describes a system exhibiting diffusive motion where $D_m$ represents the diffusion coefficient.

Given the intrinsically heterogeneous nature of such systems, multiple replicates of the same composition are often examined, increasing the already heavy computational costs that are typical of DDM analysis. %Here we speed up computation by only computing a fraction of the image structure function.
To demonstrate this, we re-analyze six replicates of the active actin data set from \cite{lee2021myosin} to show that despite  computing the Fourier transformation of the intensity difference at only 25 selected $\Delta t$'s, which significantly reduces the computational cost,   DDM-UQ  extracts the information that is as accurate as  the DDM approach that was originally employed. In detail, the full $D_o(q,\Delta t)$ is computed directly from the image stacks by DDM, from which a subset of values, $D_o(q,\Delta t)$ are pre-selected to obtain the predictive distribution by DDM-UQ. 

This example specifically compares the DDM and DDM-UQ analysis when a parametric model with four fitting parameters are specified. When a stretched exponential model [Eq. (\ref{eq:stretched exponential})] for $f(q,\Delta t)$ with coefficients $\tau(q)$ and $\gamma(q)$, along with $A(q)$ and $B(q)$, is fit to either the full matrix $D_o(q,\Delta t)$ (DDM) or the predictive distribution $D(q,\Delta t)$ (DDM-UQ) for each $q$, the estimated image structure function $D_e(q,\Delta t)$ is obtained.     $D(q,\Delta t)$ contains the same number of entries as $D_o(q,\Delta t)$; it is reconstructed using the mean of values sampled from the predictive distribution $D(q,\Delta t)$ [300 instances at every ($q$, $\Delta t)$]. The purpose of this step is to extract coefficients $\tau(q)$ and $\gamma(q)$ relevant to the underlying physical process.
 Here for DDM, we use the conventional approach by minimizing the square error loss between the model and the $D(q,\Delta t)$, which is used as the loss function, whereas for DDM-UQ, we minimize the weighted square loss where the weights are calculated from the inverse variance of the predictive samples from GPR.

The difference between the estimated quantity $D_e(q,\Delta t)$ from fitting stretched exponential model, and observed quantity $D_o(q,\Delta t)$ is evaluated by the N-RMSE:
\begin{multline}
    \mbox{N-RMSE}=\\
    \frac{\sqrt{\frac{1}{n_{q}n_{\Delta t}}\sum_{\Delta t \in \Delta \mathcal T}\sum_{q\in \mathcal Q}(\langle \tilde D_o(q,\Delta t)\rangle- \langle    \tilde D_e(q,\Delta t)\rangle )^2 }}{\tilde \sigma_{D}},
    \label{equ:nrmse_actin}
\end{multline}
 where $\tilde D_o(q,t)$ and $\tilde D_e(q,\Delta t)$ are the logarithm of the observed and estimated image structure function by different approaches, and $\tilde \sigma_{D}$ is the logarithm of sample standard deviation;  $\Delta \mathcal T$  and $\mathcal Q$ are the sets of $\Delta t$ and $q$ available for comparison, respectively.

\begin{table}[t]
\caption{N-RMSE of active actin networks.}
\centering
%\begin{tabularx}{\textwidth}{lX}
\begin{tabular}{c|cc|cc}
 \toprule
   & \multicolumn{2}{c|}{\it $<$70\%  Plateau}& \multicolumn{2}{c}{\it Full Data}\\
% \hline
Sample ID & DDM  & DDM-UQ & DDM & DDM-UQ\\
 \midrule
1 & 0.0576 & 0.0417 & 0.0700 & 0.0507 \\   
2 & 0.0641 & 0.0689 & 0.0882 & 0.0948 \\   
3 & 0.0406 & 0.0265 & 0.0573 & 0.0373 \\   
4 & 0.0759 & 0.0699 & 0.0835 & 0.0769 \\   
5 & 0.0462 & 0.0285 & 0.0576 & 0.0355 \\   
6 & 0.0588 & 0.0317 & 0.0724 & 0.0390 \\   
 \bottomrule
\end{tabular}
\label{table: GL RMSE comparison}
%\end{tabularx}
\end{table}

Figure \ref{fig:GL}(a) shows reconstructed $D(q,\Delta t)$ (dashed lines) from fitting the full observation (solid line) using DDM as well as a reconstructed $D(q,\Delta t)$ obtained by resampling with only a fraction of design points (black dots) using DDM-UQ. There is general agreement between the two approaches at all $q$ values, with some differences observed at short times because DDM-UQ weighs more heavily data at small $\Delta t$, as the predictive variance is small at these regions.
%since the number of pixels used to obtain $f(q,\Delta t)$ at small $\Delta t$ is large.
 
Fitting via a weighted least squares minimization approach is more robust for estimating the noise term $B$. One $B(q)$ is estimated to be very close to $0$ ($2.13\times 10^{-8}$) in DDM, whereas it is estimated to be  around $3.3\times 10^4$ in DDM-UQ. The significant underestimation of the noise by DDM explains the large deviation of the fit at this wave vector shown in Fig. \ref{fig:GL}(a). This example illustrates the importance of estimating the noise parameters accurately, and that fitting a parametric model of $f(q,\Delta t)$ without addressing the uncertainty by the least squared estimator can be unreliable in estimating the noise parameter. 

In Fig. \ref{fig:GL}(b) we show the same fits of the stretched exponential model  used to reconstruct $f(q,\Delta t)$ at different $q$'s using the full observed image structure function $D_o(q,\Delta t)$ obtained by the DDM (red lines) and DDM-UQ analysis (blue lines). Note that DDM-UQ only uses observations $D_o(q,\Delta t)$ at selected $q$ and $\Delta t$, but it performs equally well even for unobserved $q$ values for which there is no observation for any $\Delta t$ [see Fig. \ref{fig:GL}(a), bottom curve]. 

The differences between the estimated and observed values of $D_o(q,\Delta t)$ for DDM and DDM-UQ are quantified by the N-RMSE in Eq. (\ref{equ:nrmse_actin}) as shown in Table \ref{table: GL RMSE comparison}. Since $D(q,\Delta t)$ at large $\Delta t$ contains fewer independent samples and shows large fluctuations, it is informative to compare the accuracy of the fit only up to a threshold value, chosen here to be 70\% of the $\Delta t$ values before the plateau is reached. In all cases, DDM-UQ outperforms DDM in more closely approximating the $D_o(q,\Delta t)$ when comparing the truncated data set, and in almost all cases, the fit of DDM-UQ which has much lower computational cost,  is comparable to that of DDM, as shown in Table \ref{table: GL RMSE comparison}. Thus, DDM-UQ significantly accelerates the analysis without any observable sacrifice in accuracy with respect to post-processing of the data such as model fitting.

\begin{figure}
    \centering
    \includegraphics[width=0.35\textwidth]{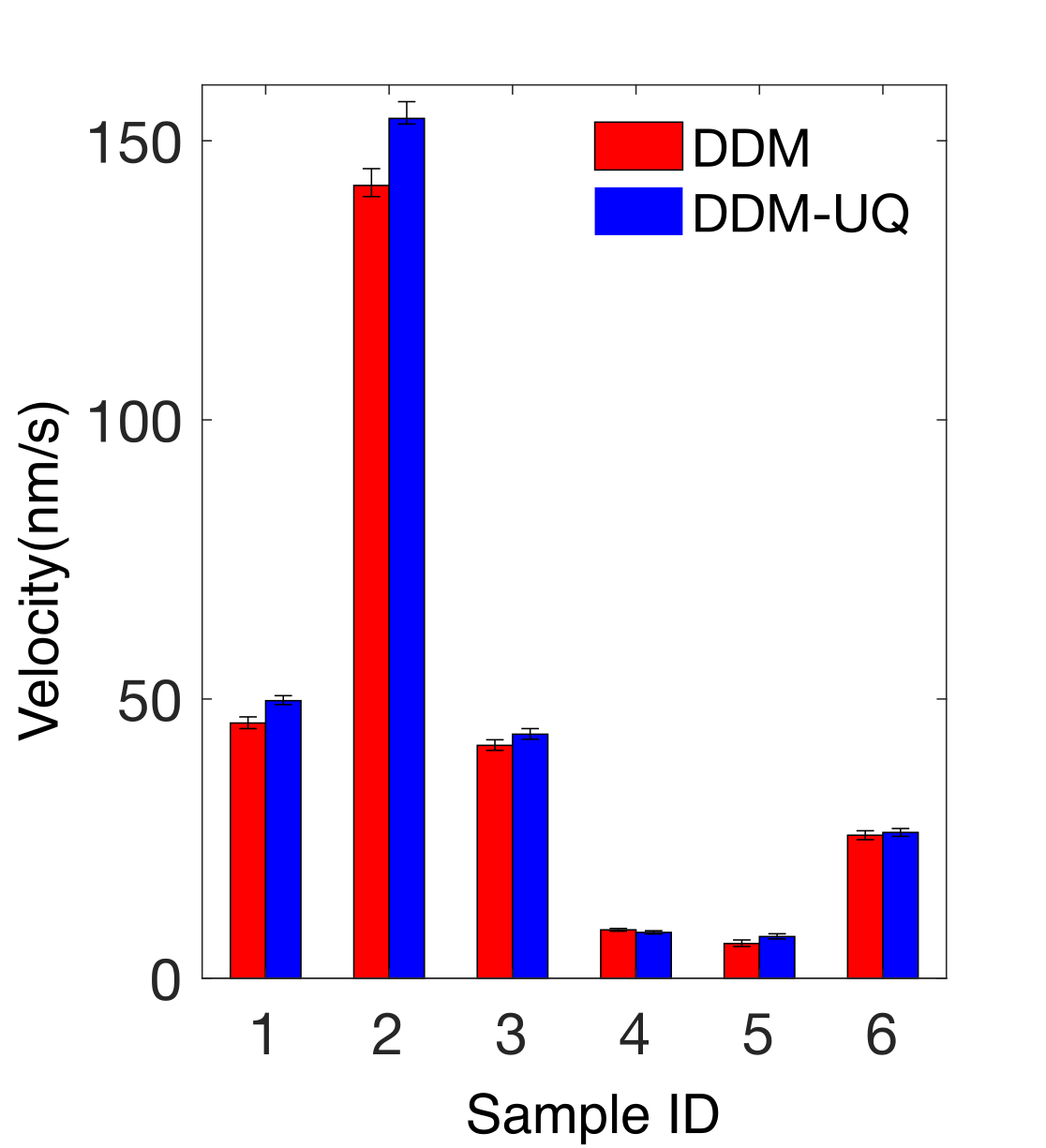}
    \caption{Velocity estimates and confidence intervals extracted from fitting the stretched exponential model to either the full observations (DDM) or to reconstructed $D(q,\Delta t)$ from selected design points (DDM-UQ). The error bars denote 95\% confidence intervals.}
    \label{fig: GL velocity MLE CI}
\end{figure}

As described in Eq. (\ref{eq:stretched exponential}), $f(q,\Delta t)$ contains the fit parameters $\gamma (q)$ and $\tau(q)$. Following \cite{lee2021myosin}, we fit the data to a linear model $\tau(q)= \frac{1}{vq}$ and extract the characteristic velocities of the active actin mixture [Fig. \ref{fig:GL}(c)]. The maximum likelihood estimator (MLE) and the confidence interval (CI) for the velocities from different replicates are tabulated in Fig. \ref{fig: GL velocity MLE CI}. Importantly, DDM-UQ with its limited observations largely recovers similar characteristic velocities and confidence intervals as those obtained using the full matrix, paving the way for high-throughput analysis of the dynamic properties of complex biomaterial systems.

\section{Concluding remarks}

DDM can be applied to a structurally evolving image stack to obtain the image structure function and intermediate scattering function. It provides an  aggregated measure of dynamics, potentially offering higher accuracy in extracting physical quantities than using real-space data alone.  
While the theoretical framework of DDM is well established, to our knowledge,  this work represents an  exploration into propagating the uncertainty associated with measurement noise, and analyzing the effects of noise in parameter estimation through mathematical and numerical analysis.  

 Based on error propagation in estimating the image structure function, we derived the mean and variance of the noise term $B$, leading to more accurate estimation of the ISF and MSD at small $\Delta t$.

 Moreover, we showed that only a small subset of $D_o(q,\Delta t)$ (around 0.5\%-5\%) at selected $q$ and $\Delta t$ need to be computed, and when they are used in a GPR model, it is possible to obtain the predictive median and samples in the image structure function at unobserved inputs and subsequent quantities of interest could be robustly predicted. Both simulations and experiments were presented to demonstrate that our method has virtually no loss of information, while reducing the computational 
time by  25-120 times. The combined improvements offered by the error propagation and predictive median from GPR results in more robust estimation of the intermediate scattering function and mean squared displacements. Through the comparisons made here between DDM-UQ, other formulations of DDM, and MPT, we highlight the need for accurate noise estimation in the analysis and interpretation of DDM experiments.  

We anticipate that these results will enable many new applications of DDM to complex biomaterial and soft material systems \cite{fricks2009time,lysy2016model}. With the potential to carry out real-time analysis via down-sampling, the proposed method can be extended to map out an entire phase space of material composition or physicochemical conditions in a high-throughput manner. This increased performance also places new demands on the general applicability of the algorithm, for instance, to provide meaningful analysis of stiffer materials that do not fully decorrelate as quickly as a more fluid-like samples, as well as heterogeneous samples, through analysis of sub-populations that demonstrate distinct features.   Future extensions of DDM-UQ analysis should include reducing the selection bias by properly weighting the ISF by the inverse variance of the noise,  which should provide a more reliable and fully automated estimation of physical quantities at larger lag times. Another potential direction is to derive a more robust estimator of the imaging noise $\sigma_0^2$ from the image stack that can contain less bias to some challenging experimental scenarios (summarized in Table  \ref{tab:potential_bias_B}).  
These directions will be pursued in future work. 

\section*{Acknowledgements}
This work was supported by the BioPACIFIC Materials Innovation Platform of the National Science Foundation under Award No. DMR-1933487 (NSF BioPACIFIC MIP), with partial support by  the Materials Research Science and Engineering Center (MRSEC) Program of the National Science Foundation under Award No. DMR-1720256 (IRG-3).  MG acknowledges partial support from the National Science foundation under Award No. DMS-2053423.  MEH acknowledges partial support from the National Science Foundation under Award No. CBET-1729108.  This work used computational facilities purchased with funds from the National Science Foundation (Grant No. CNS-1725797) and administered by the Center for Scientific Computing (CSC). The CSC is supported by the California NanoSystems Institute and MRSEC (NSF Grant No. DMR-1720256) at UC Santa Barbara. We thank G. Lee and R. Robertson-Anderson from University of San Diego for providing the active actin data set. 

\bibliography{References_2020}% Produces the bibliography via BibTeX.
\section*{Appendices}
\setcounter{figure}{0}
\renewcommand{\thefigure}{A\arabic{figure}}
\section*{Appendix A}

\begin{proof}[Proof of Eq.  (\ref{equ:representation_squared_intensity_fourier}) and Eq. (\ref{equ:I_o_fourier_squared_expection})]

The observed difference of the intensity at two time points $(t+\Delta t)$ and $t$ can be described as 
\[ \Delta I_o(\mathbf x,t, \Delta t)=\Delta I(\mathbf x,t,\Delta t)+ \Delta \epsilon(\mathbf x,t,\Delta t), \]
where $\Delta I(\mathbf x,t,\Delta t)=  I(\mathbf x,t+\Delta t)- I(\mathbf x,t)$ and $\Delta \epsilon(\mathbf x,t,\Delta t)=\epsilon(\mathbf x,t+\Delta t)-\epsilon(\mathbf x,t) $. We denote the minimum time interval by $\Delta t_{min}=1$ and $\Delta t=l \Delta t_{min}=l$, where $l$ is a positive integer smaller than $T$.  
 
We apply 2D discrete Fourier transformations on $\Delta I_o(\mathbf x,t,\Delta t)$ and obtain

\begin{align*}
&|\Delta \hat{I}_o(\mathbf q,t,\Delta t)|^2\\
=&\frac{1}{N^2}\left|\sum^{N-1}_{x_1=0}\sum^{N-1}_{x_2=0} \Delta I_o(\mathbf x,t,\Delta t) \exp\left(- \frac{i 2\pi \mathbf x^T \mathbf q}{N} \right) \right|^2 \\
=&\frac{1}{N^2}\left\{\sum^{N-1}_{x_1=0}\sum^{N-1}_{x_2=0} \Delta I_o(\mathbf x,t,\Delta t)\cos\left( \frac{2\pi\mathbf x^T\mathbf  q}{N}\right)\right\}^2 \\ 
& \quad  +\frac{1}{N^2}\left\{\sum^{N-1}_{x_1=0}\sum^{N-1}_{x_2=0} \Delta I_o(\mathbf x,t,\Delta t) \sin\left( \frac{2\pi\mathbf x^T\mathbf  q}{N}\right) \right\}^2\\
:=&{\Delta \hat I^2_{o,1}(\mathbf q,t,\Delta t)+\Delta  \hat I^2_{o,2}(\mathbf q,t,\Delta t)} \\
\end{align*}
where $\hat I_{o,1}(\mathbf q,t,\Delta t)$ and $\hat I_{o,2}(\mathbf q,t,\Delta t)$ are independent from each other by the orthogonality of the Fourier basis with
\begin{align*}
\E[\hat I_{o,1}(\mathbf q,t,\Delta t)]&=\frac{1}{N}\sum^{N-1}_{x_1=0}\sum^{N-1}_{x_2=0} \Delta I(\mathbf x,t,\Delta t) \cos\left( \frac{2\pi\mathbf x^T\mathbf  q}{N}\right),\\
\E[\hat I_{o,2}(\mathbf q,t,\Delta t)]&=\frac{1}{N}\sum^{N-1}_{x_1=0}\sum^{N-1}_{x_2=0} \Delta I(\mathbf x,t,\Delta t) \sin\left( \frac{2\pi\mathbf x^T\mathbf  q}{N}\right),\\
\V[\hat I_{o,1}(\mathbf q,t,\Delta t)]&=\V[\hat I_{o,2}(\mathbf q,t,\Delta t)]=\sigma^2_0.
\end{align*}

Furthermore 
\begin{align*}
&|\Delta \hat I_o(\mathbf q,t,\Delta t)|^2 \\
=&|\Delta \hat I({\bf q},t,\Delta t)|^2 +2\Delta \hat I({\bf q},t,\Delta t)\Delta \hat \epsilon(q, t,\Delta t) + |\Delta \hat \epsilon(q, t,\Delta t)|^2,
%&= \sum^{N-1}_{x_1=0}\sum^{N-1}_{x_2=0} \Delta I_o(\mathbf x,t,\Delta t) \left\{\cos\left(2\pi \left(\frac{x_1 q_1}{N}+\frac{x_2 q_2}{N}\right)\right)-i\sin\left(2\pi \left(\frac{x_1 q_1}{N}+\frac{x_2 q_2}{N}\right)\right) \right\}
\end{align*}
where
\begin{align}
|\Delta \hat I({\bf q},t,\Delta t)|^2&=\Delta \hat I^2_1+\Delta \hat I^2_2 , \label{equ:I_q_fourier} \\
\Delta \hat I({\bf q},t,\Delta t)\Delta \hat \epsilon(q, t,\Delta t)&= \Delta \hat I_1 \Delta \hat \epsilon_1 +\Delta \hat  I_2 \Delta \hat \epsilon_2, \label{equ:I_q_epsilon_fourier}\\
|\Delta \hat \epsilon(q, t,\Delta t)|^2&=\Delta \hat\epsilon^2_{1}+\Delta \hat\epsilon^2_{2}, \label{equ:I_epsilon_fourier}
\end{align}
with 
\begin{align*}
   \Delta \hat I_1&=\frac{1}{N}\sum^{N-1}_{x_1=0}\sum^{N-1}_{x_2=0} \Delta I(\mathbf x,t,\Delta t)\cos\left( \frac{2\pi\mathbf x^T\mathbf  q}{N}\right),  \\
    \Delta \hat \epsilon_{1}&=\frac{1}{N}\sum^{N-1}_{x_1=0}\sum^{N-1}_{x_2=0} \Delta \epsilon(\mathbf x,t,\Delta t)\cos\left( \frac{2\pi\mathbf x^T\mathbf  q}{N}\right),   \\
      \Delta \hat   I_2&=\frac{1}{N}\sum^{N-1}_{x_1=0}\sum^{N-1}_{x_2=0} \Delta I(\mathbf x,t,\Delta t)\sin\left( \frac{2\pi\mathbf x^T\mathbf  q}{N}\right),  \\
    \Delta \hat \epsilon_{2}&=\frac{1}{N}\sum^{N-1}_{x_1=0}\sum^{N-1}_{x_2=0} \Delta \epsilon(\mathbf x,t,\Delta t)\sin\left( \frac{2\pi\mathbf x^T\mathbf  q}{N}\right).  
\end{align*}
The expected value of $\E[|\hat I_o({\bf q},t,\Delta t)|^2]$ can be verified using properties of the Fourier basis. 
\end{proof}

%\begin{proof}[Proof of Theorem \ref{thm:D_q_t}]
\begin{proof}[Proof of Eq. (\ref{equ:D_o_q_delta_t}) - Eq. (\ref{equ:V_D_q_delta_t})]
The observations of image structure function can be obtained through computing ensemble average of the observed intensity 
\begin{align*}
D_o( q, \Delta t)&=\langle |\Delta \hat{I}_o ({\bf q}, t, \Delta t)|^2 \rangle\\
&=\langle |\Delta \hat{I} ({\bf q}, t, \Delta t)|^2 \rangle +\langle |\hat{\epsilon}({\bf q}, t, \Delta t)|^2 \rangle\\ 
&\quad  +2 \langle \Delta \hat{I} ({\bf q}, t, \Delta t) \Delta \hat{\epsilon}({\bf q}, t, \Delta t)\rangle
\end{align*}
with expected value and variance:
\begin{align*}
&\E[D_o({ q}, \Delta t)]\\
&=\E[\langle |\Delta \hat{I} ({\bf q}, t, \Delta t)|^2 \rangle]+\E[\langle |\hat{\epsilon}({\bf q}, t, \Delta t)|^2 \rangle]\\
&\quad +2\E[\langle \Delta \hat{I} ({\bf q}, t, \Delta t) \Delta \hat{\epsilon}({\bf q}, t, \Delta t)\rangle]\\
&=\langle |\Delta \hat{I} ({\bf q}, t, \Delta t)|^2 \rangle  +\frac{1}{n_{\Delta t} n_q}\sum_{t \in \mathcal S_{\Delta t}}\sum_{(q_1,q_2) \in \mathcal S_q}\E[\Delta \hat{\epsilon}_1^2+\Delta \hat{\epsilon}_2^2]\\
& \quad + \frac{2}{n_{\Delta t} n_q}\sum_{t \in \mathcal S_{\Delta t}}\sum_{(q_1,q_2) \in \mathcal S_q}\Delta \hat{I} ({\bf q}, t, \Delta t) \E[\Delta \hat{\epsilon}({\bf q}, t, \Delta t)]\\
&=D( q, \Delta t)+\V(\Delta \hat{\epsilon}_1)+\V(\Delta \hat{\epsilon}_2)\\
&=D( q, \Delta t)+2\sigma_0^2
%\label{equ:E_Dqt}
\end{align*}
where $\langle \cdot \rangle$ denotes averaging over available time points for each $
\Delta t$, and $(q_1, q_2) \in \mathcal S_q$ with $ \mathcal S_q= \{ (q_1,q_2):q_1^2+q_2^2 =q^2 \}$, $n_q = \# \mathcal S_q$, $n_{\Delta t}= T-\Delta t$, and 
\begin{align*}
    &\V[D_o( q, \Delta t)]\\
    &=\V[\langle |\Delta \hat{I}_o ({\bf q}, t, \Delta t)|^2 \rangle]\\
    &=\V \Big [\frac{1}{n_q n_{\Delta t} }\sum_{(q_1,q_2) \in \mathcal S_q}\sum_{t \in \mathcal S_{\Delta t}} |\Delta \hat{I}_o ({\bf q}, t, \Delta t)|^2\Big] \quad\quad\quad\quad
\end{align*}
%It's not hard to verify that 
The variance of the average of intensity over available time points for each $\Delta t$ is
\begin{align*}
    &\V \big[\frac{1}{n_{\Delta t}}\sum_{t \in \mathcal S_{\Delta t}}|\Delta \hat{I}_o ({\bf q},t, \Delta t)|^2\big]\\
    &=\V \Big [\frac{1}{n_{\Delta t}}\sum_{t \in \mathcal S_{\Delta t}}\big (\Delta \hat I^2_{o,1}(\mathbf q,t,\Delta t)+\Delta  \hat I^2_{o,2}(\mathbf q,t,\Delta t)\big)\Big ]\\
    &=\frac{1}{n_{\Delta t}^2}\sum_{t \in \mathcal S_{\Delta t}} \V[\Delta \hat I^2_{o,1}(\mathbf q,t,\Delta t)]\\
    & \quad +\frac{1}{n_{\Delta t}^2}\sum_{t \in \mathcal S_{\Delta t}}\V[\Delta \hat I^2_{o,2}(\mathbf q,t,\Delta t)]\\
    & \quad  +\frac{1}{n_{\Delta t}^2}\sum_{t_1 \neq t_2} Cov\big( \Delta \hat I^2_{o,1}(\mathbf q,t_1,\Delta t), \Delta \hat I^2_{o,1}(\mathbf q,t_2,\Delta t)\big)\\
    & \quad +\frac{1}{n_{\Delta t}^2}\sum_{t_1 \neq t_2} Cov\big( \Delta \hat I^2_{o,2}(\mathbf q,t_1,\Delta t), \Delta \hat I^2_{o,2}(\mathbf q,t_2,\Delta t)\big)
\end{align*}
where the first two terms can be computed as 
\begin{align*}
    &\frac{1}{n_{\Delta t}^2}\sum_{t \in \mathcal S_{\Delta t}} \Big( \V[\Delta \hat I^2_{o,1}(\mathbf q,t,\Delta t)] + \V[\Delta \hat I^2_{o,2}(\mathbf q,t,\Delta t)]\Big)\\
    &=\frac{1}{n_{\Delta t}^2}\sum_{t=1}^{n_{\Delta t}}\Big(\V[(\Delta \hat{I}_1+\Delta \hat{\epsilon}_1)^2]+\V[(\Delta \hat{I}_2+\Delta \hat{\epsilon}_2)^2]\Big)\\
    &=\frac{1}{n_{\Delta t}^2}\sum_{t=1}^{n_{\Delta t}} \Big( \V[2\Delta \hat{I}_1 \Delta \hat{\epsilon}_1+\Delta \hat{\epsilon}_1^2] +\V[2\Delta \hat{I}_2 \Delta \hat{\epsilon}_1+\Delta \hat{\epsilon}_2^2]\Big)\\
    &=\frac{1}{n_{\Delta t}^2}\sum_{t=1}^{n_{\Delta t}} \Big( 4 \sigma^2_0(\Delta \hat{I}_1^2+\Delta \hat{I}_2^2)+ \V(\Delta \hat{\epsilon}^2_1)+\V(\Delta \hat{\epsilon}^2_2)\Big)\\
    &=\frac{1}{n_{\Delta t}^2}\sum_{t=1}^{n_{\Delta t}} \Big( 4 \sigma^2_0|\Delta \hat{I} ({\bf q}, t, \Delta t)|^2+ 4\sigma_0^4\Big),
\end{align*}
and for $\Delta t = 1,2,\dots,\lfloor (T-1)/2\rfloor$, the last two terms follows:
\begin{align*}
 &\sum_{t_1 \neq t_2} \Cov\big( \Delta \hat I^2_{o,1}(\mathbf q,t_1,\Delta t), \Delta \hat I^2_{o,1}(\mathbf q,t_2,\Delta t)\big)\\
 &=2\sum_{t=1}^{T - 2\Delta t} \Cov\big( \Delta \hat I^2_{o,1}(\mathbf q,t,\Delta t), \Delta \hat I^2_{o,1}(\mathbf q,t+\Delta t,\Delta t)\big)\\
 &=2\sum_{t=1}^{T - 2\Delta t} \Cov \big( (\Delta \hat{I}_{1}(\mathbf q,t,\Delta t)+ \hat{\epsilon}_{1,t+\Delta t}-\hat{\epsilon}_{1,t})^2,\\
 &  \quad (\Delta \hat{I}_1(\mathbf q,t+\Delta t,\Delta t)+\hat{\epsilon}_{1,t+2\Delta t}-\hat{\epsilon}_{1,t+\Delta t})^2\big)\\
 &=2\sum_{t=1}^{T - 2\Delta t} \Cov\big(\E[(\Delta \hat{I}_{1}(\mathbf q,t,\Delta t)+ \hat{\epsilon}_{1,t+\Delta t}-\hat{\epsilon}_{1,t,\Delta t})^2],\\
 & \quad \E[(\Delta \hat{I}_1(\mathbf q,t+\Delta t,\Delta t)+\hat{\epsilon}_{1,t+2\Delta t}-\hat{\epsilon}_{1,t+\Delta t})^2]  |\hat{\epsilon}_{1,t+\Delta t}\big)\\
 & +2\sum_{t=1}^{T - 2\Delta t}\E\big[ \Cov((\Delta \hat{I}_{1}(\mathbf q,t,\Delta t)+ \hat{\epsilon}_{1,t+\Delta t}-\hat{\epsilon}_{1,t})^2,\\
 & \quad (\Delta \hat{I}_1(\mathbf q,t+\Delta t,\Delta t)+\hat{\epsilon}_{1,t+2\Delta t}-\hat{\epsilon}_{1,t+\Delta t})^2 |\hat{\epsilon}_{1,t+\Delta t})\big] \\
 &=2\sum_{t=1}^{T - 2\Delta t} \Big( \Cov(\hat{\epsilon}_{1,t+\Delta t}^2,\hat{\epsilon}_{1,t+\Delta t}^2)-4 \times \\
 & \quad \Delta \hat{I}_{1}(\mathbf q,t,\Delta t)\Delta \hat{I}_{1}(\mathbf q,t+\Delta t,\Delta t) \Cov(\hat{\epsilon}_{1,t+\Delta t},\hat{\epsilon}_{1,t+\Delta t}) \Big)\\
 &=2\sum_{t=1}^{T - 2\Delta t} \Big( \frac{\sigma_0^4}{2}-2\sigma_0^2 \Delta \hat{I}_{1}(\mathbf q,t,\Delta t)\Delta \hat{I}_{1}(\mathbf q,t+\Delta t,\Delta t)\Big),\\
\end{align*}
with 
\begin{align*}
    \hat{\epsilon}_{1,t}&=\frac{1}{N}\sum^{N-1}_{x_1=0}\sum^{N-1}_{x_2=0}\epsilon(\mathbf x, t)\cos\left( \frac{2\pi\mathbf x^T\mathbf  q}{N}\right) \\
    \hat{\epsilon}_{2,t}&=\frac{1}{N}\sum^{N-1}_{x_1=0}\sum^{N-1}_{x_2=0}\epsilon(\mathbf x, t)\sin\left( \frac{2\pi\mathbf x^T\mathbf  q}{N}\right). 
\end{align*}
Similarly, for $\Delta t = 1,2,\dots,\lfloor (T-1)/2\rfloor$ 
\begin{align*}
    &\sum_{t_1 \neq t_2} \Cov\big( \Delta \hat I^2_{o,2}(\mathbf q,t_1,\Delta t), \Delta \hat I^2_{o,2}(\mathbf q,t_2,\Delta t)\big)\\
    &=2\sum_{t=1}^{T - 2\Delta t} \Big( \frac{\sigma_0^4}{2}-2\sigma_0^2 \Delta \hat{I}_{2}(\mathbf q,t,\Delta t)\Delta \hat{I}_{2}(\mathbf q,t+\Delta t,\Delta t)\Big)
\end{align*}

For general $\Delta t>\lfloor (T-1)/2\rfloor$:
\begin{align*}
    &\sum_{t_1\neq t_2} \Cov\big( \Delta \hat I^2_{o,1}(\mathbf q,t_1,\Delta t), \Delta \hat I^2_{o,1}(\mathbf q,t_2,\Delta t)\big)\\
    &=\sum_{t_1 \neq t_2} \Cov\big( \Delta \hat I^2_{o,2}(\mathbf q,t_1,\Delta t), \Delta \hat I^2_{o,2}(\mathbf q,t_2,\Delta t)\big)\\
    &=0
\end{align*}

Combining the variance and covariance expressions developed above, the variance of the average of intensity is
\begin{align*}
    &\V \big[\frac{1}{n_{\Delta t}}\sum_{t \in \mathcal S_{\Delta t}}|\Delta \hat{I}_o ({\bf q},t, \Delta t)|^2\big]\\
    &= \frac{2\sigma_0^2}{n_{\Delta t}} \Big(2 \sigma_0^2 +\frac{2\sum^{n_{\Delta t}}_{t=1}|\Delta \hat{I}(\mathbf q,t,\Delta t)|^2}{n_{\Delta t}} \\
    & \quad \quad \quad +\max(0,T-2\Delta t)\big(\frac{\sigma_0^2}{n_{\Delta t}}-\frac{2 S_{q_1,q_2,\Delta t}}{n_{\Delta t}(T-2\Delta t)} \big)\Big)
\end{align*}
with
\begin{align}
    S_{q_1,q_2,\Delta t}&=\sum_{t=1}^{T - 2\Delta t}\Big(\Delta \hat{I}_1(\mathbf q,t,\Delta t)\Delta \hat{I}_1(\mathbf q,t+\Delta t,\Delta t) \nonumber \bigg.\\
    & \quad \quad +\Delta \hat{I}_2(\mathbf q,t,\Delta t)\Delta \hat{I}_2(\mathbf q,t+\Delta t,\Delta t)\Big) 
    \label{eq:S_q1_q2}
\end{align}

Finally, we have
\begin{align*}
    &\V[D_o( q, \Delta t)]\\
    &=\V \Big [\frac{1}{n_q}\sum_{(q_1,q_2) \in \mathcal S_q} \frac{1}{ n_{\Delta t} }\sum_{t \in \mathcal S_{\Delta t}} |\Delta \hat{I}_o ({\bf q}, t, \Delta t)|^2\Big]\\
    &=\frac{1}{n_q^2}\sum_{(q_1,q_2) \in \mathcal S_q}\frac{2\sigma_0^2}{n_{\Delta t}} \Big(2 \sigma_0^2 +\frac{2\sum^{n_{\Delta t}}_{t=1}|\Delta \hat{I}(\mathbf q,t,\Delta t)|^2}{n_{\Delta t}}\\
    & \quad \quad \quad +\max(0,T-2\Delta t)\big(\frac{\sigma_0^2}{n_{\Delta t}}-\frac{2 S_{q_1,q_2,\Delta t}}{n_{\Delta t}(T-2\Delta t)} \big)\Big)\\
    &=\frac{2\sigma_0^2}{n_q n_{\Delta t}}\Big(2\sigma_0^2+2D(q,\Delta t)\\
    &\quad \quad \quad +\max(0,T-2\Delta t)\big(\frac{\sigma_0^2}{n_{\Delta t}}-\frac{2 S_{q,\Delta t}}{(T-2\Delta t)n_{\Delta t}n_q} \big)\Big)
\end{align*}
with 
\begin{align}
    S_{q,\Delta t}=\sum_{(q_1,q_2):q_1^2+q_2^2=q^2}S_{q_1,q_2,\Delta t}
    \label{eq:S_q}
\end{align}

%Additional proof for the variance of the average of intensity over available time points: 

\end{proof}

\section*{Appendix B Parameter estimation in Gaussian process regression}

Let $\mathbf {\tilde D}_o=(\tilde D_o(\bm \theta_1),...,\tilde D_o(\bm \theta_n))^T$ denote the $n$ observations. The parameters in the Gaussian process contain mean parameter $m$, variance parameter $ \sigma^2$ and inverse range parameters $\bm \beta=(\beta_1,\beta_2)$ in the kernel function. Conditional on  $\bm \beta$ and the regularization parameter $\lambda$, the maximum likelihood estimator of the mean parameter is $  { m}_{est}=(\mathbf 1^T_n \mathbf {\tilde R}^{-1} \mathbf 1^T_n)^{-1}\mathbf 1^T_n \mathbf {\tilde R}^{-1} \mathbf {\tilde D}_o$, with $\mathbf {\tilde R}=\mathbf R+n\lambda \mathbf I_n$ 
and $ {\sigma}^2_{est}=S^2/n$ with $S^2=(\mathbf {\tilde D}_o- \mathbf 1_n  \hat {\mathbf m} )^T \mathbf {\tilde R}^{-1}(\mathbf {\tilde D}_o- \mathbf 1_n  \hat {\mathbf m} )$. Plugging   $( m_{est},  \sigma^2_{est})$, the profile likelihood of  parameters $(\bm \beta, \lambda)$ in the covariance function follows $\mathcal L(\bm \beta,\lambda )\propto |{\mathbf K}|^{-\frac{1}{2}} (S^2)^{-\frac{n}{2}} $. Since  no closed formed expression of the maximum likelihood estimator for the range and regularization parameters is available, one often numerically maximizes the profile likelihood to obtain the estimates of $(\bm \beta,\lambda)$. When the sample size is small, the MLE can be unstable and marginal posterior mode estimation with robust parametrization is often used \cite{gu2018robust}. We implemented the parameter estimation and predictions of GPR by the ``RobustGaSP" package available in R and MATLAB \cite{gu2018robustgasp}.

\section*{Appendix C Detection of estimator bias}

\begin{figure}
    \centering
    \includegraphics[width=0.38\textwidth]{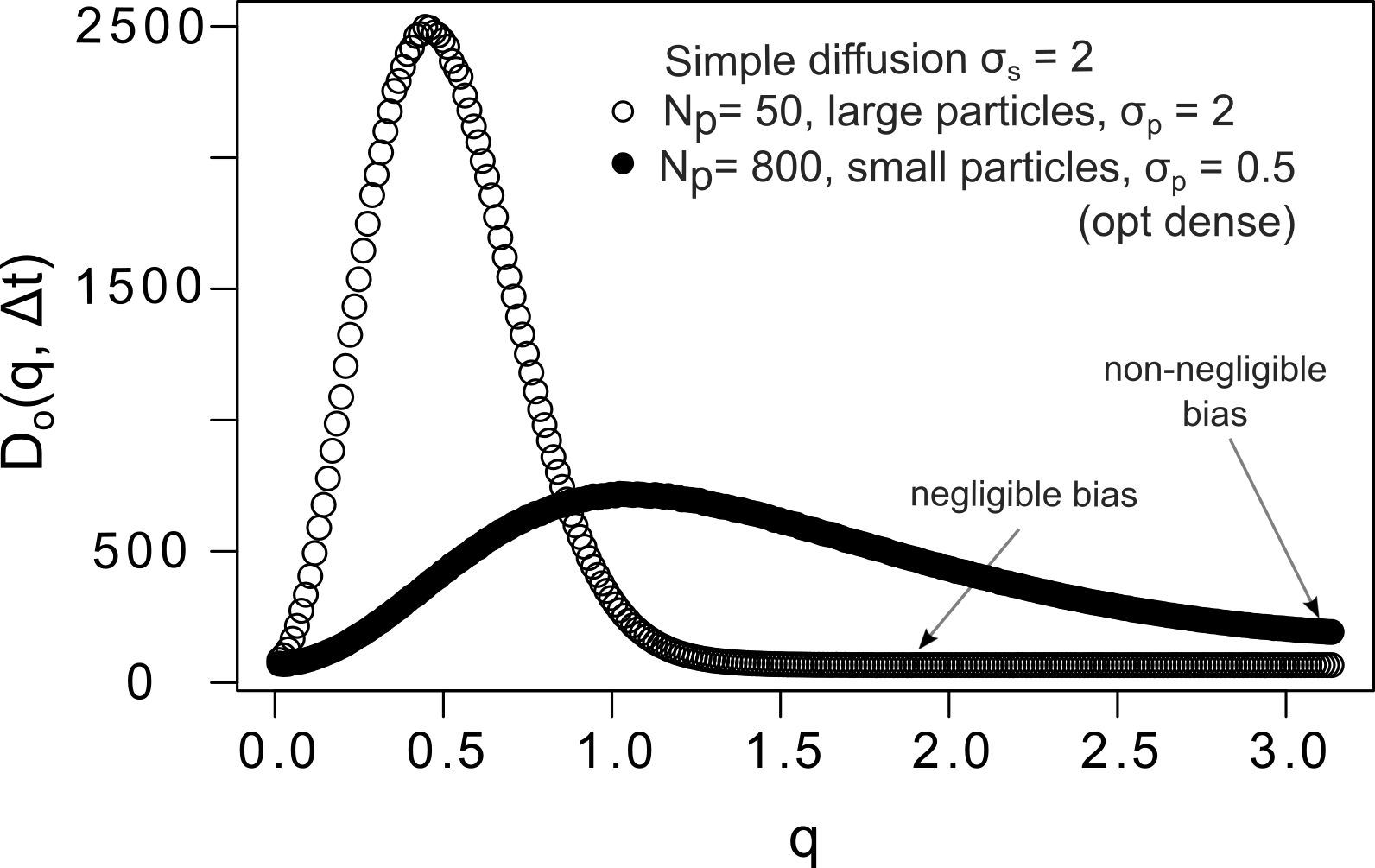}
    \caption{ A close look at the estimators of noise $B_{est}$ in the first and second scenarios of simple diffusion with identical underlying dynamics $\sigma_s$ =2, above. The difference lies in that the second case includes 16$\times$ as many particles while particle radius decreases by 4$\times$ (filled circles). For the range of $q$ probed, $D_o(q,\Delta t)$ does not decay to zero at $q_{max}$.}
    \label{fig:DetectBias}
\end{figure}

The overestimation by a series of previously used methods can be detected by plotting $D_o( q,\Delta t)$ over all $q$ values at one $\Delta t$ (see Fig. \ref{fig:DetectBias}). If at high $q$, $D_o( q,\Delta t)$'s rate of change slows, then this implies that $A(q)(1-f(q,\Delta t))$ is close to zero,  and the bias in estimating the noise term  by the second to the fourth estimator in Table \ref{tab:potential_bias_B} is negligible. However, if $D_o( q,\Delta t)$ decreases (even slightly) as the value $q$ increases, then the bias of the estimator is non-negligible.

\begin{figure}
    \centering
    \includegraphics[width=0.4\textwidth]{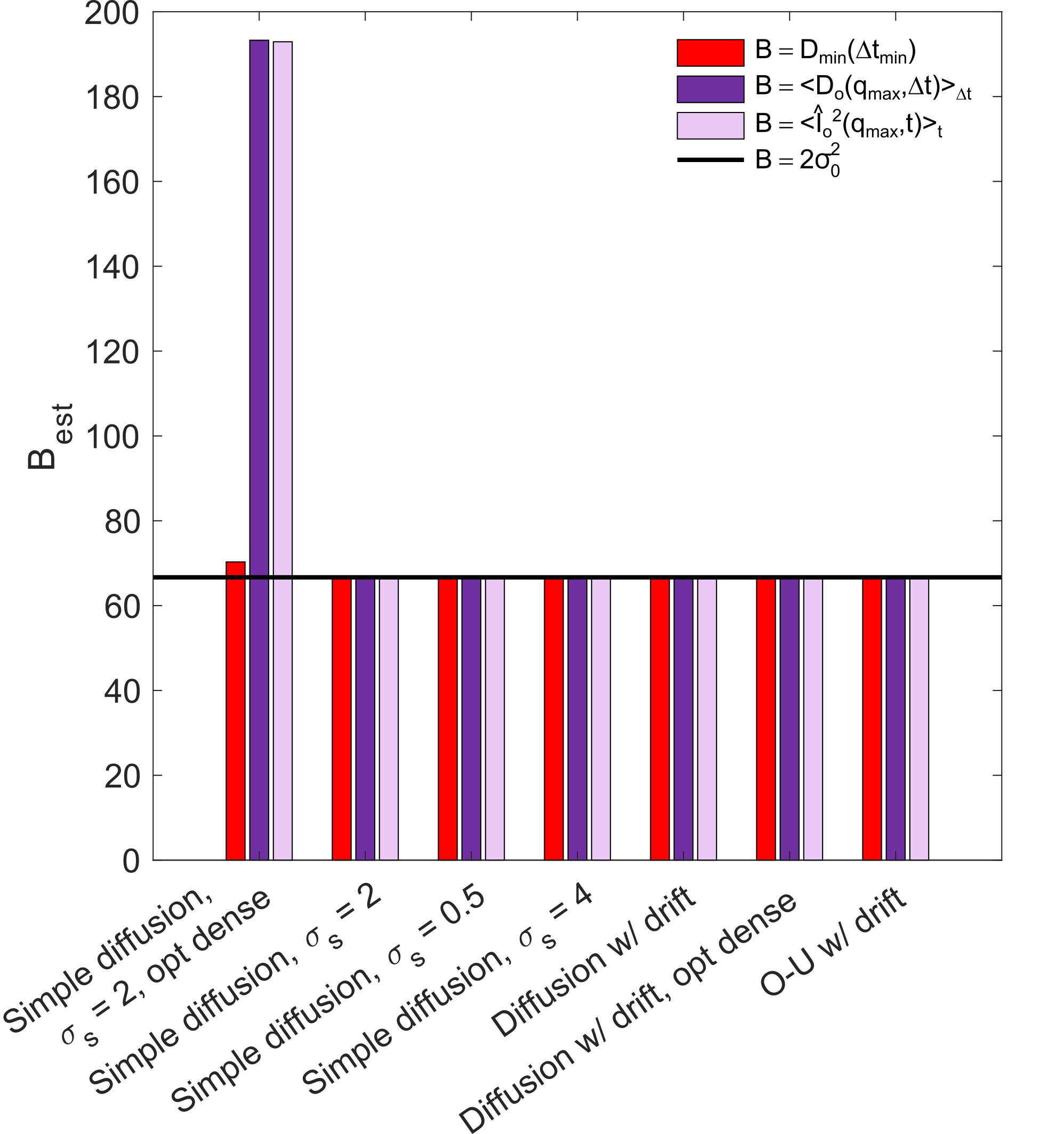}
    \caption{Comparison of different estimators (denoted by different color bars) for all simulation scenarios. Note that the true noise $2\sigma_0^2$ is kept constant in all simulations and denoted by the thick black line. }
    \label{fig:EstB}
\end{figure}

\section*{Appendix D Derivation of mean squared displacement}
Here we derive the MSD for each of the  different scenarios explored in simulation.
%W.l.o.g., assume we observe particle movement $\mathbf x(t) \in \mathbb  \Delta r^2$ at $t=$ 
First, the simulated particles in  Sections IIIB and IIIC all undergo Brownian motion. Without loss of generality, we may assume the variance of $\Delta x_{i,j}(t)$ is $\sigma^2_s$. For any $ \mathbf x(t), \mathbf x(t+\Delta t) \in  \mathbb R^2$, the MSD can be simply computed by 
\begin{align*}
&\E[ ( x_{ij}(t+\Delta t)-  x_{ij}(t))^2] \\
=&\V\left[ x_{ij}(t+\Delta t)-  x_{ij}(t)\right]+\left\{\E[  x_{ij}(t+\Delta t)-  x(t)]\right\}^2 \\
=&\V\left[ \sum^{\Delta t-1}_{k=0} \Delta x_{ij}(t+k)\right]+0=\sigma^2_s \Delta t,
\end{align*}
for any $j=1,...,p$ and $i=1,2$. 
Since particles move isotropically in a 2D space, the MSD is $2 \sigma^2_s \Delta t$. 

In simulated scenarios presented in Section IIID, similarly we can split the  MSD into two terms. Noting 
$\E[  x_{ij}(t+\Delta t)-  x_{ij}(t)]=\mu_D {\Delta t}$ and the process is isotropic, the MSD is  $2 \sigma^2_s \Delta t+ 2\mu^2_D \Delta t^2$, 

In simulated scenarios presented in Section IIIE, note that when $t=t_1$, $x_{ij}(t_1) \sim N(x_{ij}(t_0), \sigma^2_D)$. For any $t>t_1$, it is not hard to show 
\begin{align*}
\mathbb E[x_{1j}(t)]&=(t-1)\mu_D cos(\theta_j)+x_{1j}(t_0)\\
\mathbb E[x_{2j}(t)]&=(t-1)\mu_D sin(\theta_j)+x_{2j}(t_0), 
\end{align*}
and consequently 
\begin{equation}
   \mathbb E[\sum^2_{i=1}(x_{ij}(t+\Delta t)- x_{ij}(t))^2]= \mu^2_D\Delta t^2. 
   \label{equ:E_diff}
\end{equation}
It is easy to verify $\mathbb V[x_{ij}(t)]=\sigma^2_s$. Thus we have 
\begin{align}
  &\mathbb V[x_{ij}(t+\Delta t)- x_{ij}(t)] \nonumber\\
  =&  \mathbb V[x_{ij}(t+\Delta t)]+\mathbb V[x_{ij}(t)]- 2\mbox{Cov}(x_{ij}(t+\Delta t), x_{ij}(t)) \nonumber \\
  =&2\sigma^2_s-2\mbox{Cov}(\rho^{\Delta t}x_{ij}(t), x_{ij}(t))\nonumber \\
  =& 2\sigma^2_s-2\sigma^2_s \rho^{\Delta t}
  \label{equ:V_diff}
\end{align}
Since the process is on a two dimensional space, combining (\ref{equ:E_diff}) and (\ref{equ:V_diff}), the MSD is 
$4\sigma^2_s-4\sigma^2_s\rho^{\Delta t}+\mu_D^{2}\Delta t^{2}$. 

\section*{Appendix E Notations}
The notations used in this paper are listed in Table \ref{table: table of notations}.
\begin{table*}[htb!]
\caption{Table of notations.}
\begin{tabularx}{0.9\textwidth}{@{}XX@{}}
\toprule
  {\it Brackets, overheads \& superscripts} \\
  $\tilde{~}$ & logarithmically transformed variables \\
  $\hat{~}$ & Fourier transformed functions \\
  $<\dots>_i$     & ensemble average with respect to variable $i$\\
  $|\dots|$         & modulus of complex numbers\\
  $\dots_o$ & observed quantities\\
  $\dots_*$ & unknown variables \\   
  $\dots_e$ &estimated quantities \\
  $\Delta$ & difference of functions or variables\\
  $\dots^T$     & transpose\\
  \smallskip
  {\it Variables} \\
  ${\bf x}$   & coordinates in real-space \\
  ${\bf q}$   & coordinates in Fourier transformed (reciprocal) space\\
  $q$         & radius of coordinates in Fourier transformed space\\
  $t$         & real time in experiments \\
  $t_{min}$   & starting time  in experiments \\
    $t_{max}$   & ending time  in experiments \\
  $T$         & the total number of time points in experiments \\
  $\Delta t$  & lag time\\
  $\epsilon$  & noise in the image intensity\\
  $\sigma_o^2$  & variance of the noise in the image intensity\\
  $n_p$         & number of particles\\
  $n_{\Delta t}$         & number of $\Delta t$'s when lag time is $\Delta t$\\
  $n_{q}$         & number of coordinates with radius $q$\\
  $\tau$      & range parameter in a stretched exponential model\\
  $\gamma$      & roughness parameter in a stretched exponential model\\
  $\mathcal S$  & sets of variables\\
  $\Delta \mathcal T$ & sets of available $\Delta t$'s\\
  $\mathcal Q$ & sets of available $q$'s\\
  
  \smallskip
  {\it Particle Dynamics Simulations} \\
  $\sigma_s$    & diffusive step size \\
  $\mu_D$       & drift velocity \\
  $D_m$         & diffusion coefficient \\
  $I_p({\bf x},t)$ & particle intensity function \\
  $I_c$         & particle center pixel intensity used in $I_p({\bf x},t)$\\
  $I_b({\bf x},t)$ & background intensity function \\
  \smallskip
  {\it Gaussian Process Regression}  \\
  $m$       & mean parameter\\
  $\sigma^2$    & variance parameter\\
  $\bm \beta$      & inverse range parameters\\
$\lambda$        &    regularization  parameter \\
  $\epsilon_\theta$ & noise in regression\\
  $\bf {D_o}$   & vector of observed image structure function\\
  \smallskip
  {\it Constants} \\
  $k_B$         & Boltzmann constant \\
  $T_a$         & absolute temperature \\
  \smallskip
  
  {\it Functions \& Operators} \\
  $I({\bf x},t)$ & image intensity function \\
  $D({\bf q},\Delta t)$ & image structure function\\
  $D_o({\bf q},\Delta t)$ & observed image structure function (which contains noise)\\
  $D_e({\bf q},\Delta t)$ & estimated image structure function fit by parameterized model\\
  $f({\bf q},\Delta t)$ & intermediate scattering  function \\
  $\langle   \Delta r^2(\Delta t)\rangle$ & mean squared displacement \\
  $\mathcal F(\dots)$  & operator of the 2D discrete Fourier transform\\
  $\mathcal L(\dots)$  & likelihood function\\
  $\mathcal N$ & normal distribution\\
  $\V(\dots)$      & variance of variables\\
  $\E(\dots)$      & expected value of variables\\
  \bottomrule
\end{tabularx}
\label{table: table of notations}
\end{table*}

\end{document}